\documentclass[utf8]{frontiersSCNS} 

\usepackage{url,hyperref,lineno,microtype,subcaption}
\usepackage[onehalfspacing]{setspace}
\usepackage{aasmacros}
\usepackage{tabularx}

\def\keyFont{\fontsize{8}{11}\helveticabold }
\def\firstAuthorLast{Husser {et~al.}} 
\def\Authors{Tim-Oliver Husser\,$^{1,*}$, Frederic V.\ Hessman\,$^{1}$, Sven Martens\,$^{1}$, Tilman Masur\,$^{1}$, Karl Royen\,$^{1}$, and Sebastian Sch\"afer$^{1}$}
\def\Address{$^{1}$Institute for Astrophysics and Geophysics, G\"ottingen University, G\"ottingen, Germany}
\def\corrAuthor{Tim-Oliver Husser}

\def\corrEmail{thusser@uni-goettingen.de}

\begin{document}
\onecolumn
\firstpage{1}

\title[pyobs]{pyobs - An observatory control system for robotic telescopes} 

\author[\firstAuthorLast ]{\Authors} 
\address{} 
\correspondance{} 

\extraAuth{}

\maketitle

\begin{abstract}

\section{}
We present a Python-based framework for the complete operation of a robotic telescope observatory. It provides out-of-the-box support for many popular camera types while other hardware like telescopes, domes, and weather stations can easily be added via a thin abstraction layer to existing code. Common functionality like focusing, acquisition, auto-guiding, sky-flat acquisition, and pipeline calibration are ready for use. A remote-control interface, a ``mastermind'' for truly robotic operations as well as an interface to the Las Cumbres Observatory observation portal is included. The whole system is fully configurable and easily extendable. We are currently running pyobs successfully on three different types of telescopes, of which one is a siderostat for observing the Sun. pyobs uses open standards and open software wherever possible and is itself freely available.

\tiny
 \keyFont{ \section{Keywords:} methods: observational, telescopes, techniques: image processing, techniques: photometric, techniques: spectroscopic} 
\end{abstract}

\section{Introduction}
At the turn of the millennium, a major change was starting to take place in observational astronomy: at first unnoticed by most astronomers, many new telescopes were refitted or newly built for remote or even fully autonomous observations. While there were lots of discussions and even a few robotic telescopes in the early 90s \citep[see, e.g.,][]{1992ASPC...34...67P,1992ASPC...34..193A}, their number grew significantly in the decades thereafter.

While most of the very early robotic telescopes simply monitored known variable stars \citep[e.g.,][]{1995AJ....110.2926H,1997A&AS..125...11S}, those that followed were designed to permit very rapid follow-up of gamma-ray bursts -- e.g., ROTSE-III \citep{2003PASP..115..132A}, REM \citep{2003MmSAI..74..304A}, and BOOTES \citep{2004AN....325..679C} -- or search for exoplanets \citep[e.g., with SuperWASP,][]{2003ASPC..294..405S}, or to survey galaxies for supernovae \citep[e.g.][]{2001ASPC..246..121F,2007A&AT...26...79L}.  As automation became easier and pipeline software more powerful, it was possible to survey automatically for any transients or moving Solar System objects, e.g. with the Intermediate Palomar Transient Factory \citep[iPTF,][]{2009PASP..121.1395L}, which later was refitted to become the Zwicky Transient Facility \citep[ZTF,][]{2019PASP..131f8003B,Riddle2018}. About the same time, Las Cumbres Observatory (LCO) started building a whole network of robotic telescopes \citep{2008AN....329..269H}, now one of the largest in the world.

Robotic telescopes can be used for many things -- from the automated performance of a heterogeneous list of independent observations to the dedicated performance of a particular scientific project. The unique science that can be done with robotic telescopes almost exclusively concerns transients, i.e.\ changes over time on the sky of any kind. The ultimate source for such targets in the near future will be the Legacy Survey of Space and Time (LSST) at the Vera C. Rubin Observatory, an $8.4\,\textrm{m}$ telescope designed for surveying the sky for any kind of transients \citep{2019ApJ...873..111I}. When the LSST starts operating in late 2022, a legion of other robotic telescopes will begin doing follow-up observations on the detected transients.

All truly robotic telescopes require a wide palette of hardware -- e.g. computer-controlled telescopes, cameras, filter wheels, enclosures, and weather stations -- and software for autonomous operation of the entire system. While this software can be rather basic (to avoid the fully unwarranted word "simple") for surveys that just do the same thing over and over again, it gets immensely more complicated for all-purpose telescopes. Unfortunately, this software almost never gets published or placed in a form which is useful for another project, mostly for the reason that it is very specific to the hardware and the science case at hand.

Luckily, there is some software available that tries to be applicable to many different hardware devices and kinds of observations. Especially popular with amateur astronomers is the Windows COM-based ASCOM system\footnote{\url{https://ascom-standards.org}}, which defines generic interfaces for different kinds of devices and can be used by several client applications. A couple of years ago, a HTTP REST based interface called Alpaca was released, which allows the use of ASCOM in Unix-like systems as well.
Additional powerful software like ACP from DC3.com can be used to help automate operations within an ASCOM network.
A very different system but with the same basic philosophy and breadth of support is the Instrument Neutral Distributed Interface (INDI).\footnote{\url{https://indilib.org}}, which was designed for network transparency from the beginning and can be used from any system and programmed in any language, although the core libraries are written in C++.
The 2nd version of the Remote Telescope System (RTS2 for short; \cite{2004AIPC..727..753K}) is widely used in a variety of mostly scientific projects. It provides a complete framework -- including a back-end database -- and is designed for fully autonomous operations. RTS2 is written in C++ and runs on Linux only.

For our own MONET telescopes \citep[see also Section~\ref{sec:telescopes}]{2004AN....325..533H} we first successfully used the robotic control software developed for their twins, the STELLA telescopes \citep{2006AN....327..792G,2012ASInC...7..247G} on Tenerife, operated by the AIP in Potsdam, which was thankfully made available to us by our colleagues there. While the system itself is written in Java, over time we started to implement some functionality using more familiar Python scripts. These scripts grew and at some point became pyobs, a fully functional observation control system for robotic telescopes on its own.  There was originally no other good reason for developing pyobs than this; without pyobs and starting from scratch, we probably would have chosen INDI. However, pyobs has now grown to a level where it is just as powerful as INDI, RTS2, or ASCOM: it is highly flexible, uses open standards, and is programmed in the language most commonly used by astronomers.  Indeed, it stands on the shoulders of giants that are the many amazing open source Python projects used in computer science and astronomy. In this paper we will present its architecture and the basic functionality.

We strongly believe in acknowledging the work other people put into publicly available (open-source) software, and thus, references for all the third party software projects used in pyobs are listed in the Acknowledgments. All pyobs packages themselves are published as open-source under the MIT license at GitHub\footnote{\url{https://github.com/pyobs}}, and its documentation is also available online.\footnote{\url{https://www.pyobs.org}}

\section{Architecture}
The astronomical community has spent the last two decades migrating from diverse programming languages like IDL, FORTRAN, or C/C++ to a common denominator, which turned out to be Python. As a result, today we have powerful scientific libraries available like NumPy, SciPy, and AstroPy. Following this progress, Python was an easy pick as the language of choice for a new Observatory Control System (OCS).

Nevertheless, Python has some drawbacks for a large project like this, with the "global interpreter lock" (GIL) being the most significant. The GIL is a multi-threading lock (or "mutex") that can only be acquired by one thread at a time. So, although Python supports the creation and running of multiple threads, they never run in parallel. The only way to achieve true parallelism is to use multi-processing, so a decision was made to run pyobs in multiple processes, i.e.\ one process per block of functionality, which, in pyobs terminology, is called a "module". A module can be everything from a controller for an actual hardware device to routines for, e.g., an auto-focus series. With the OCS being split up into multiple processes, the communication between them became one of the most important parts of pyobs.

\subsection{Communication}
Instead of inventing our own protocol for communication, we decided to use XMPP \citep{RFC3920}, an XML-based chat protocol. With it being mainly used for instant messaging (e.g.\ by Jabber, WhatsApp, Zoom, Jitsi, and others), it naturally supports multi-user chat, i.e.\ sending messages to multiple users. But due to its wide variety of extensions (XMPP Extension Protocol, XEP), it also supports remote procedure calls (RPC, calling methods on another client), and a feature called auto-discovery, which allows one client to determine the capabilities of another.

The use of XMPP also frees us from writing and maintaining our own server software, since there are multiple industrial-grade servers available, like ejabberd\footnote{\url{https://www.ejabberd.im}} and Openfire\footnote{\url{https://www.igniterealtime.org/projects/openfire/}}. They can run with tens of thousands of users, compared to maybe a few dozen pyobs clients in a typical observatory. Although, admittedly, pyobs sends more messages than even the most ambitious teenager in WhatsApp.

While we use the Python package Slixmpp for pyobs itself, there are also XMPP libraries available for all major programming languages\footnote{see, e.g., \url{https://xmpp.org/software/libraries/}}. Therefore the "py" (for "Python") in "pyobs" refers only to the core package, but extension modules can be written in any language that supports XMPP.

\begin{figure}[t]
  \begin{center}
    \includegraphics[width=\textwidth]{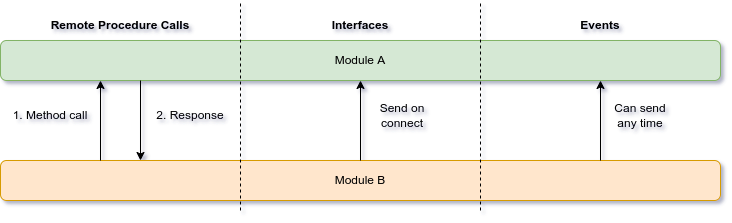}
  \end{center}
  \caption{The three pillars of communication in pyobs. On the left, remote procedure calls are actively called on another module. The list of interfaces, in the middle, is automatically retrieved after the connection to the XMPP server has been established. And events can be sent at any time, as shown on the right. Only those events can be handled in a module that it had registered before.}
  \label{fig:comm}
\end{figure}
As Fig.~\ref{fig:comm} shows, the communication in pyobs is based on three pillars (remote procedure calls, interfaces, events), which all will be discussed in more detail in the following.

\subsubsection{Remote procedure calls}
All methods within a module that are derived from an interface (see below) can be called remotely. The easiest way to do so, is to get a \texttt{Proxy} object for another module from pyobs. These objects mimic the behavior of the original module and therefore any of their methods can be called directly as if they were local.

For instance, a camera module might implement this method:
\begin{verbatim}
    async def set_exposure_time(
        self, 
        exposure_time: float, 
        **kwargs: Any
    ) -> None:
\end{verbatim}
For another module, calling this method is a simple as:
\begin{verbatim}
    camera = await self.proxy(name_of_camera_module)
    await camera.set_exposure_time(2.0)
\end{verbatim}

While some methods should usually return immediately (e.g.\ requesting a position), some might take a longer time (e.g.\ exposing an image or moving a telescope). For the caller of a method it would be good to have an estimate for the call duration in order to avoid waiting forever in case of an error. To achieve this, pyobs extends the XEP-0009 extension for RPCs with a timeout mechanism: all methods can define a time after which they should be finished. This time is sent back to the caller immediately after a method is called. If this waiting time is exceeded, a timeout exception is raised and the caller can decide what to do about this. If a method takes longer than 10\,seconds, it should be decorated with the \texttt{@timeout} decorator, which defines the maximum duration:
\begin{verbatim}
    @timeout(1200)
    async def move_radec(
        self, 
        ra: float, 
        dec: float, 
        **kwargs: Any
    ) -> None:
\end{verbatim}
Calling a method like this works the same way as before, although it now raises an exception only after $1200\,\mathrm{s}$, compared to $10\,\mathrm{s}$ for un-decorated methods.

\subsubsection{Interfaces}
\label{sec:interfaces}
The basis for all RPCs in pyobs are the interfaces in \texttt{pyobs.interfaces}, which describe methods that a module must implement in order to provide a given functionality. For instance, all telescope modules should implement the \texttt{ITelescope} interface. While not defining any methods on its own, it inherits the two methods \texttt{move\_radec} and \texttt{get\_radec} from \texttt{IPointingRaDec} (shortened for clarity):
\begin{verbatim}
    class IPointingRaDec(Interface):
        @abstractmethod
        async def move_radec(
            self, 
            ra: float, 
            dec: float, 
        ) -> None:
            ...

        @abstractmethod
        async def get_radec(self, ) -> Tuple[float, float]:
            ...

    class ITelescope(IPointingRaDec):
        ...
\end{verbatim}
Therefore, in order to be a valid \texttt{ITelescope}, a module must implement these methods.

All interfaces implemented by a module are published via XMPP's auto-discovery extension, so all other modules can easily determine what functionality is available from a given module. This allows for easy construction of \texttt{Proxy} objects for RPC. Furthermore, it is extremely simple for a module to find all other modules that implement a given interface. A good example for this are the interfaces \texttt{IFitsHeaderBefore} and \texttt{IFitsHeaderAfter}. When a camera starts a new exposure, we usually want to collect FITS headers from different modules. Instead of having this list pre-defined, the camera can just request all modules that implement these interfaces and call their respective methods before and after the exposure:
\begin{verbatim}
    clients = await self.comm.clients_with_interface(
        IFitsHeaderBefore
    )
    for client in clients:
        proxy = await self.proxy(client, IFitsHeaderBefore)
        headers[client] = await proxy.get_fits_header_before()
\end{verbatim}
This way, we can easily add a new module to the system that simply provides new headers for new FITS files (e.g.\ with weather data).

\begin{figure}[t]
  \begin{center}
    \includegraphics[width=\textwidth]{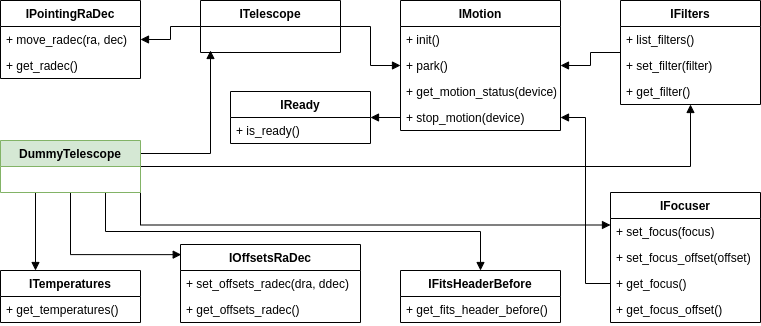}
  \end{center}
  \caption{Part of the interface inheritance for \texttt{DummyTelescope} (on the left in green), a simulated telescope that accepts RA/Dec coordinates and offsets and has a filter wheel, a focus unit, and some temperature sensors. All methods available for remote calls are defined in the interfaces. \texttt{ITelescope} does not define any method of its own, but is just a collection of other interfaces and can be used as a device definition, i.e.\ "this is a telescope".}
  \label{fig:interfaces}
\end{figure}
As an example, Fig.~\ref{fig:interfaces} shows parts of the inheritance for \texttt{DummyTelescope}, a simulated telescope that can be used for testing.

\subsubsection{Events}
While RPC is an active process of communicating with other modules, there is also a passive one, which is reacting to events. Each module can define types of events that itself creates and that it wants to receive from other modules.

For instance, a camera might want to declare that it can send events, when a new image has been taken
\begin{verbatim}
    await self.comm.register_event(
        NewImageEvent
    )
\end{verbatim}
and can actually send those events:
\begin{verbatim}
    await self.comm.send_event(
        NewImageEvent(filename, image_type)
    )
\end{verbatim}
while another module might want to receive those events and handle them in a callback method:
\begin{verbatim}
    await self.comm.register_event(
        NewImageEvent, 
        self.on_new_image
    )
    
    [...]
    
    async def on_new_image(
        self, 
        event: Event, 
        sender: str
    ) -> bool:
        print(event)
\end{verbatim}

The events can be chained by sending new events within a handler method. As an example, events on new images from a camera could be handled by an image pipeline, which in turn sends events that are handled by a module that measures seeing on the reduced images.

\subsection{Configuration}
pyobs gets its high flexibility from configuration files in YAML format. The most simple configuration consists of only a single line like:

\begin{verbatim}
    class: pyobs.modules.test.StandAlone
\end{verbatim}

When running this configuration via \texttt{pyobs config.yaml} from the command line, a new module is created from the given class and started. The class to use is given by its full package name, the same as one would use to import it in a Python shell. Therefore, its definition could be anywhere within the Python path and not just in the pyobs package.

The example in the documentation is a little longer:
\begin{verbatim}
    class: pyobs.modules.test.StandAlone
    message: Hello world
    interval: 10
\end{verbatim}
Comparing this with the signature of the constructor of the given class:
\begin{verbatim}
    class StandAlone(Module):
    def __init__(
        self, 
        message: str = "Hello world", 
        interval: int = 10, 
        **kwargs: Any
    ):
\end{verbatim}
This makes it clear that all items in the configuration are simply forwarded directly to the constructor of the given class. pyobs goes even a step further and allows many parameters to be either an object or a configuration dictionary (mostly given in a YAML file as in the example above), describing an object of the same type. For instance, every module class has also a parameter \texttt{comm} (derived from pyobs' \texttt{Object} class) for defining its method for communication with other modules, given as this:
\begin{verbatim}
    comm: Optional[Union[Comm, Dict[str, Any]]] = None
\end{verbatim}
So this parameter accepts both a \texttt{Comm} object directly or a description thereof. A valid configuration file could therefore look like this:
\begin{verbatim}
    class: pyobs.modules.test.StandAlone
    comm:
        class: pyobs.comm.slixmpp.XmppComm
        jid: test@example.com
        password: topsecret
\end{verbatim}
Looking at the constructor of given class \texttt{XmppComm} explains the given parameters (shortened for clarity):
\begin{verbatim}
    class XmppComm(Comm):
        def __init__(
            self,
            jid: Optional[str] = None,
            password: str = "",
        ):
\end{verbatim}
Note that this makes it possible to replace the whole communication system via XMPP with another method by just implementing a new class derived from \texttt{Comm}. In an environment, in which it is impossible to run an XMPP server, this could simply be replaced by, e.g., direct socket communication or HTTP REST.

A similar configuration style is used for names of remote modules, which are called within a module. Here is the constructor of the default class for taking an auto-focus series (shortened):
\begin{verbatim}
    class AutoFocusSeries(Module, IAutoFocus):
        def __init__(
            self,
            focuser: Union[str, IFocuser],
            camera: Union[str, IImageGrabber],
        ):
\end{verbatim}
The class needs two remote modules to work, a camera for taking the images and a focus unit with which it can change the actual focus value. Both are defined to accept either a string or an object implementing the interface that is actually required. While for testing, it might be easier to pass an actual object, at the observatory we usually just set the name of the other module. From this name, a proxy object is being created, which is checked for implementing the given interface. Therefore, in production, a configuration for a focus series might look like this:
\begin{verbatim}
    class: pyobs.modules.focus.AutoFocusSeries
    camera: fli230
    focuser: telescope
\end{verbatim}
Note that all this behavior is completely up to the class that you want to use. So it must implement the flexibility to accept both an object and a description or a remote name. This should be the case for all modules from the core package and the additional packages.

If possible, the configuration even allows changing the core behavior of a module. Coming back to the \texttt{AutoFocusSeries} class from above, this class itself only defines the functionality for taking a series of images at different focus values. The actual analysis of the images and the calculation of the final best focus is delegated to an object of type \texttt{FocusSeries} as defined as a parameter in the constructor:
\begin{verbatim}
    series: Union[Dict[str, Any], FocusSeries]
\end{verbatim}
The default implementation in pyobs (\texttt{ProjectionFocusSeries} in \texttt{utils.focusseries}) collapses the images along their x and y axes, respectively, and calculates moments to get a rough size of the stars. The final best focus is calculated using a hyperbola fit to the series of focus and size data. But, given that this class is explicitly specified in the configuration file, it can easily be changed to another (custom) implementation that derives from \texttt{FocusSeries}.

A module might want to make some configuration settings changeable during runtime. This can be handled via the \texttt{IConfig} interface, which is implemented by default by all modules and calls internal methods of the form \texttt{\_set\_config\_<name>} (if exists) for changing the given variable \texttt{<name>}.

\subsection{asyncio}
Given the already mentioned problems with multi-threading in Python, it is only logical to rethink the use of threads in pyobs in the first place. Most modules in pyobs do one thing most and foremost: waiting. Waiting for a command to execute, waiting for an exposure to finish, waiting for the dome to move into position. However, in pyobs many things still need to be run concurrently, e.g.\ a module should still be accepting commands while moving a telescope. Luckily, Python introduced a new way of handling concurrency in version 3.5 and improved it steadily in the years thereafter. The new asyncio package uses a main loop and switches between tasks on request, all on a single CPU core and in a single thread. This avoids typical problems in multi-threading like deadlocks and run conditions. However, calling a blocking function in asyncio blocks all other tasks as well, so there is also an easy way for running single methods in an extra thread and waiting for it.

Functions that are running within the asyncio loop are called \textit{coroutines} and are defined with the \texttt{async} keyword, as shown before for the interfaces:
\begin{verbatim}
    class IPointingRaDec(Interface):
      async def move_radec(ra: float, dec: float):
        ...
\end{verbatim}

Coroutines can only be called directly from other coroutines and always need to be "awaited":
\begin{verbatim}
    async def test():
      await telescope.move_radec(1., 2.)
\end{verbatim}
They can also be called without actually waiting for them to finish. In those cases, a task should be created which can be awaited later:
\begin{verbatim}
    task = asyncio.create_task(telescope.move_radec(1., 2.))
    ...
    await task
\end{verbatim}
In pyobs, this is, for instance, used for requesting FITS headers from other modules before an exposure is started. The module creates tasks for requesting the headers, but only awaits them after the image has finished, in order not to delay the start of the exposure.

As mentioned above, asyncio heavily reduces the risk of multi-threading related problems. That is, because tasks never run in parallel, but are only switched when one has finished or when something is awaited. In multi-threading, parts of the code that should not be interrupted are often secured using a mutex (or lock), which is mostly unnecessary when using asyncio.

With asyncio, one just needs to be careful with long running functions that are not defined \texttt{async}, e.g.\ the readout processes of some cameras. Those method calls would block the whole module, so asyncio provides an easy way to run them in an extra thread:
\begin{verbatim}
    loop = asyncio.get_running_loop()
    data = await loop.run_in_executor(None, camera.read_out())
\end{verbatim}

Altogether, pyobs make heavy use of asyncio. For instance, all interface methods and all event handlers must be defined \texttt{async}. Switching from multi-threading to asyncio massively reduced the number of difficult-to-debug errors and made developing a lot easier.

\subsection{Virtual File System}
\label{sec:vfs}
In a simple pyobs system, all its modules might run on a single computer. In that case, a module storing a file on a local disk can be certain that another module can access it at the same location. An easy workaround for using this system with modules on different machines is to mount (e.g.\ via NFS or SMB) the required directories on both machines, but even in that case one has to be careful to mount to the same directory, otherwise filenames would not be the same on both.

This is where a virtual file system (VFS) becomes useful: if we could define a "virtual" directory that points to the correct location on all computers, the problem would be solved. pyobs provides a VFS in \texttt{pyobs.vfs} and uses it wherever files are accessed. The VFS is automatically available in all modules, although it needs to be configured. A simple VFS configuration (within the module configuration) might look like this:
\begin{verbatim}
    vfs:
      class: pyobs.vfs.VirtualFileSystem
      roots:
        temp:
          class: pyobs.vfs.LocalFile
          root: /data/images
\end{verbatim}
The VFS in pyobs uses the concept of "roots" to define where a file is actually located. In this case, one root, \texttt{temp}, is defined as a \texttt{LocalFile}, which itself has a \texttt{root} parameter, pointing to a real directory in the file system -- note that \texttt{root} here has nothing to do with the roots system in pyobs' VFS, but comes from the term "root directory".

Now, within a pyobs module with this configuration we can open a file like this:
\begin{verbatim}
    fd = self.vfs.open_file("/temp/new/image.fits", "r")
\end{verbatim}
Internally, pyobs maps the first part of the path (the root), i.e.\ \texttt{temp} in this case, to the root of the same name given in the configuration, so it actually creates a \texttt{LocalFile}. When opening the file, the path is changed accordingly to \texttt{/data/images/new/image.fits}. Following up on the example from above, now the \texttt{temp} root can point to different directories on all computers, but still the same filenames can be used on all.

Since the mounting of remote directories might not be possible in some cases, pyobs offers some more classes for file access within the VFS:
\begin{itemize}
    \item \texttt{ArchiveFile} connects to the pyobs-archive image archive (see Section~\ref{sec:pyobs-archive}). Currently only writing is permitted, i.e.\ uploading an image to the archive.
    \item \texttt{HttpFile} represents a file on a HTTP server, e.g.\ the pyobs file cache (see Section~\ref{sec:filecache}).
    \item \texttt{LocalFile} is a local file on the machine the module is running on.
    \item \texttt{MemoryFile} stores a file in memory.
    \item \texttt{SMBFile} allows access to a file on a Windows share without mounting it.
    \item \texttt{SSHFile} accesses a file on a remote machine that is accessible via SSH.
    \item \texttt{TempFile} works on a temporary file that will be deleted after being closed.
\end{itemize}

A file opened via VFS almost works like a normal file-like object in Python, with the one difference that all its methods are \texttt{async}, so they need to be awaited. pyobs also offers some convenience functions for reading and writing FITS, YAML, and CSV files in the VFS.

\begin{figure}[t]
  \begin{center}
    \includegraphics[width=\textwidth]{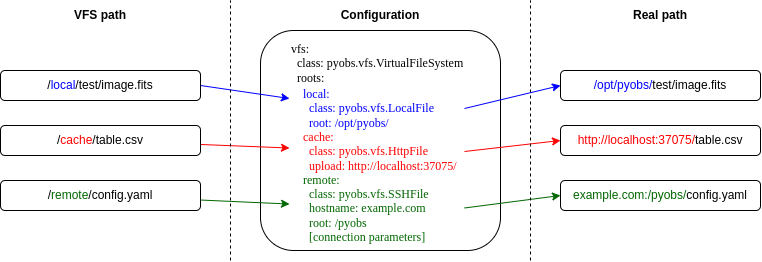}
  \end{center}
  \caption{Some examples, how a VFS path (left) maps to a real path (right) with a given configuration (middle). Note that for the \texttt{remote} root the class \texttt{SSHFile} requires more parameters for the connection, which have been omitted here for clarity.}
  \label{fig:vfs}
\end{figure}
Figure~\ref{fig:vfs} shows some examples, how a VFS path maps to a real path with a given configuration.

\subsection{Image processors and pipelines}
\label{sec:images}
With the \texttt{Image} class in \texttt{pyobs.images}, pyobs offers a class for reading and writing images that also has support for additional data like a good pixel mask, a star catalog and pixel uncertainties. It is a simple wrapper around the FITS functionality in astropy and is used within pyobs whenever images need to be passed along.

Building on this image class, pyobs has the concept of "image processors" (defined in \texttt{pyobs.images.processors}), which simply take an image, process it in some way, and then return it. Currently, these types of processors are available:
\begin{itemize}
    \item An \textbf{astrometry} processor takes an existing catalog attached to the image and tries to plate-solve it.
    \item The \textbf{detection} processors try to detect objects in the image and write a catalog.
    \item The processors in \textbf{exptime} try to estimate a good exposure time from an image, the one existing implementation is for star fields.
    \item In \textbf{offsets} are processors that calculate some kind of offsets, usually used for guiding and acquisition.
    \item The \textbf{photometry} processors perform photometry on the image (usually at positions determined using a detection step) and write/extend the catalog.
    \item There are some more \textbf{misc} processors that can, e.g., add a good pixel mask, calibrate the image, or bin it.
\end{itemize}

Since image processors take an image as input as well as returning one as output, they can easily be chained into an image pipeline. This is done by many modules for pre-processing images in some (fully customizable) way before working on them. Adding new image processors is easily done and provides a perfect way for handling images.

pyobs also offers a full (offline) image pipeline (in \texttt{utils.pipeline.Night}) that is also based on image processors, permitting the fully automatic processing of a night's images.

\subsection{Error handling}
\label{sec:error_handling}
Handling errors in a single program is sometimes difficult enough, but it can get rather complicated in a distributed system like pyobs. The basic requirement for every module is that it should handle errors on its own as well as possible (e.g., resolve errors states in hardware devices) but sometimes a calling module needs to be informed about a problem, e.g.\ if a camera does not respond to requests anymore.

Since error handling can be very specific to the problem at hand, pyobs only provides a framework for dealing with this, not a final solution. It introduces its own set of exceptions that are all derived from \texttt{PyObsError} in \texttt{pyobs.utils.exceptions}, and new exceptions can easily be added if required.

A module can call \texttt{register\_exception()} and define a callback that is called whenever a given exception is raised and a given condition is met: the function accepts a limit of how often this can happen (optionally in a given time span) before the problem is escalated. In that case, the raised exception is changed into a \texttt{SevereError}, keeping the original exception as an attribute. That means, catching one of these severe errors means that an error has occurred too often (in a given time span).

This gets more interesting in a real pyobs system with several modules. There are some cases, in which a module should stop working at all and inform other modules about this. So, e.g., the \texttt{BaseCamera}, which is the base class for all cameras it pyobs, registers an exception like this:
\begin{verbatim}
    register_exception(
      GrabImageError, 
      3, 
      timespan=600, 
      callback=self._default_remote_error_callback
    )
\end{verbatim}
This defines that after three occurrences of the \texttt{GrabImageError} exception within 600\,s, the given method should be called, which is a default implementation in \texttt{Module}. It simply logs the error and sets the module to an error state that prevents (almost) any of its methods to be invoked remotely. If another module tries to call methods anyway, it receives a \texttt{ModuleError}.

If a method is invoked remotely and an exception is raised, this exception is wrapped in a \texttt{RemoteError} with the original exception stored in an attribute. This is useful to register exceptions with the \texttt{module} parameter, which only registers an exception on a given remote module. For instance, the \texttt{FocusSeries} module uses this:
\begin{verbatim}
    register_exception(
      RemoteError, 
      3, 
      timespan=600, 
      module=camera, 
      callback=self._default_remote_error_callback
    )
\end{verbatim}
So, whenever the remote module \texttt{camera} raises too many exceptions, the \texttt{FocusSeries} module itself goes into error state, which can be cleared remotely by calling \texttt{reset\_error()} -- and of course might reappear when the exception is raised again.

Note that registering an exception always also registers parent exceptions. So if exception \texttt{B} is derived from \texttt{A}, all occurrences of \texttt{B} also count for the registered limits for \texttt{A}.

\section{Available modules}
\label{sec:modules}
In general there are two types of modules coming with pyobs: those that control actual hardware and those that do not. While the latter are part of the core package, the former are outsourced to separate packages, since they will not be required by everyone and often need special drivers to be installed. All modules can be found on the central GitHub page.

For developing your own modules, please refer to the documentation or just have a look at the existing ones as examples. There is also a simulation available that can be used for first tests. Please see the documentation for details on how to set it up.

Table~\ref{tbl:modules} lists all modules available in the core package and in external packages.

\subsection{Cameras}
pyobs knows two kinds of cameras: classic cameras (derived from the interface \texttt{ICamera}), for which one actually starts and stops an exposure, and webcam-like cameras (interface \texttt{IVideo}, which constantly provide a video (or a series of images) as output. In addition, spectrographs are also supported (interface \texttt{ISpectrograph}), which output a spectrum instead of an image -- therefore, most spectrographs would be implemented as a camera, since they return an image, from which the spectrum needs to be extracted.

In the following those camera types are listed, for which stable modules exist and are available via GitHub and PyPi. In addition, we also have modules for Andor and QHYCCD cameras, as well as normal USB webcams (via Video4Linux2), but they are all not in a publishable state. If you need one of those, please contact the author of this paper.

\subsubsection{SBIG}
The pyobs-sbig package builds on the SbigDevKit Linux driver for SBIG cameras. It is based on a Cython wrapper around that library's \texttt{CSBIGCam} and \texttt{CSBIGImg} classes. The different modules support SBIG cameras with and without filter wheel. There is a additional implementation for the STXL-6303E, due to its different gain at different binnings. Note that this special treatment of single models might be necessary for other cameras. The module has been tested on STXL-6303E, STF-402M, and STF-8300M cameras.

\subsubsection{Finger Lakes Instrumentation}
A Cython wrapper around the official \texttt{libfli}\footnote{\url{https://www.flicamera.com/downloads/FLI_SDK_Documentation.pdf}} library for FLI cameras is the core of the pyobs-fli. The module has been tested on a FLI ProLine 230.

\subsubsection{ZWO ASI}
pyobs-asi is a thin wrapper around the zwoasi package to support the cameras by ZWO ASI. It has been tested on a ZWO ASI071MC Pro.

\subsubsection{Aravis}
Aravis\footnote{\url{https://github.com/AravisProject/aravis}} is a library for Genicam cameras connected via gigabit ethernet or USB3. The module in pyobs-aravis uses a modified version of the python-aravis package for communicating with the cameras. It has been tested with several cameras from The Imaging Source\footnote{\url{https://www.theimagingsource.de}}.

\subsection{Other hardware}
While astronomical cameras are often bought off the shelf and a few brands are most common between observatories, this is mostly quite different for the other hardware in the dome -- and the dome itself. Those devices are often operated by custom controllers and need special treatment. However, if a driver of any kind exists, it is very simple to write a wrapper for it to be used within a pyobs system.

An attempt to standardize the communication between all kinds of devices has been made with ASCOM. A pyobs module for ASCOM will be described in detail below. Another of those attempts is INDI, for which we do not have a pyobs wrapper yet. Interfaces to those two standards are an easy way to add hardware to a pyobs system, for which ASCOM/INDI drivers already exist.

\subsubsection{ASCOM}
ASCOM is a standard for communicating with astronomical devices in Windows and is supported by a wide range of cameras, telescopes, domes, etc. Furthermore, there a many client applications like "The Sky" or "Stellarium" that can operate an ASCOM based system.

While pyobs can run on Windows, we made the experience that some things are a little more prone to error on that operating system -- pyobs processes sometimes quit without warning. There is a (private) pyobs package for calling ASCOM interfaces directly on Windows, but due to these problems, we never published it. However, we can provide access on request.

The restriction to Windows systems is due to the use of Windows COM as means for communication, which is not available for other operating systems. Luckily, in 2018 ASCOM presented a new interface, called Alpaca, which is based on HTTP REST requests, and therefore can also be accessed from Unix-like systems. The pyobs-alpaca package provides modules for telescopes, domes and focus units via Alpaca. However, in contrast to most other modules in the pyobs ecosystem, these ones are not meant to be used directly, but more as some kind of inspiration for an observatory specific implementation. They are not a general implementation of the ASCOM protocol, but tailored specifically for the use case of the 50cm Cassegrain telescope based at the Institute for Astrophysics and Geophysics in G\"ottingen.

\subsubsection{Pilar}
Pilar is a telescope control software from "4pi Systeme"\footnote{\url{http://www.sonobs.de/company/company.html}} based on the Open Telescope Software Interface (OpenTSI), and currently used by our MONET telescopes via the pyobs-pilar package. While the specific implementation of this module might not be of interest for most observatories, it shows an example for a socket based communication protocol wrapped in a pyobs module.

\subsection{Automating}
While the modules described so far are all built around a specific piece of hardware, there are also those that purely consist of software to automate the boring stuff.

\subsubsection{Auto-focus}
A common problem in astronomy is focusing the image on the camera sensor. In most cases this will be done by moving either a mirror (mostly the secondary) or the camera back and forth until stars appear sharp, i.e.\ with the smallest possible width. The \texttt{AutoFocusSeries} (in \texttt{pyobs.modules.focus}) module accomplishes this by taking a series of images at different focus values (i.e.\ position of M2 or camera), and tries to find an optimal focus by fitting a hyperbola through the estimated star widths in each image as a function of focus value. For this, references to a camera and a focus unit must be specified so that they can be controlled remotely.

The estimation of star sizes is fully configurable by injecting a class implementing the \texttt{FocusSeries} (in \texttt{pyobs.utils.focusseries}) interface. Our current default implementation is defined in \texttt{ProjectionFocusSeries}, which projects the image along its x and y axis, respectively, and measures moments on the resulting 1D data. Another possibility is to use a method for star detection/photometry for estimating star widths, as used in \texttt{PhotometryFocusSeries}.

While especially smaller telescope will typically work well with a constant focus value throughout the whole night, larger telescopes (with a steel structure) are constantly changing their size (and therefore the position of the perfect focus) due to temperature changes. For these cases, pyobs provides a temperature model for the focus, which is implemented in the \texttt{modules.focus.FocusModel} module and can adjust the focus continuously throughout the night. The configuration needs to specify a model function like this:
\begin{verbatim}
    model: -0.043*T1 - 0.03*T2 + 0.06*temp + 41.69
\end{verbatim}
While the value for \texttt{temp} is automatically fetched from a given weather module (see Section~\ref{ssec:utilities.weather}), those for \texttt{T1} and \texttt{T2} must also be specified in the configuration. In this case they are mirror temperatures and are supposed to be requested from the telescope module:
\begin{verbatim}
    temperatures:
        T1:
            module: telescope
            sensor: T1
        T2:
            module: telescope
            sensor: T2
\end{verbatim}
For this to work, a module named \texttt{telescope} must exist and its \texttt{get\_temperatures} method must return values for \texttt{T1} and \texttt{T2}. With these values the module now calculates new focus values at a given interval and sets them accordingly.

An \texttt{AutoFocusSeries} also sends an event, when it has successfully determined a new focus, which can be handled by the \texttt{FocusModel} automatically to optimize its temperature model. For that to work, the model function must be defined with variables that can be fitted:
\begin{verbatim}
    model: a*T1 + b*T2 + c*temp + d    
\end{verbatim}
In this case, a set of default values must also be provided:
\begin{verbatim}
    coefficients:
        a: -0.043
        b: -0.031
        c: 0.062
        d: 41.694
\end{verbatim}
If this is set up correctly, a fully robotic system can perform multiple focus series during each night (e.g.\ if there is nothing else to do) and automatically optimize the focus temperature model over time.

\subsubsection{Flat-fielding}
A task that is prone to be automated as early as possible is flat-fielding. While this is quite simple in a controlled environment with a closed dome, e.g.\ with a flat-field screen, it becomes more challenging when done on-sky during twilight. In that case, exposure times have to be adjusted continuously to obtain optimal count rates on the images.

\begin{figure}[t]
\begin{center}
\includegraphics[width=\textwidth]{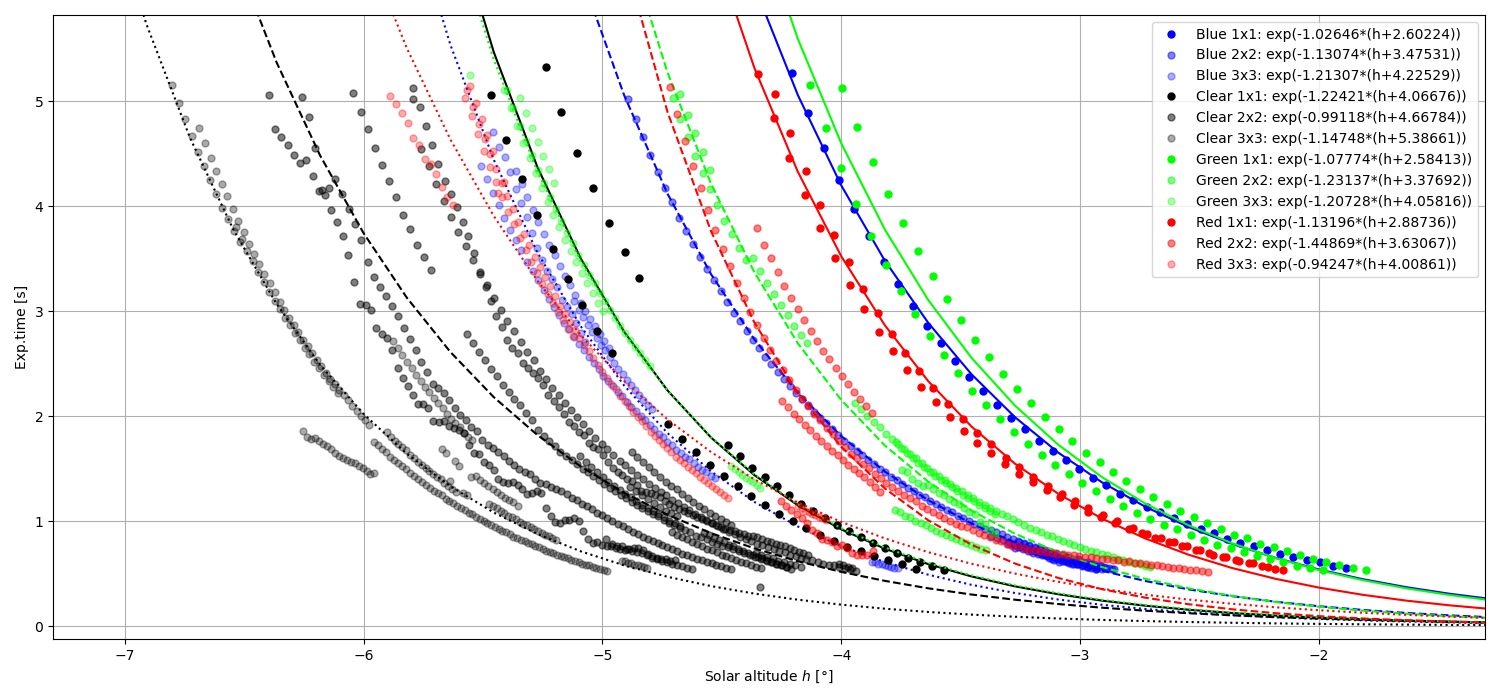}
\end{center}
\caption{An example for empirical models for flatfield exposure times. The points are optimal exposure times for getting a mean flux of 30,000 counts in the image as a function of solar altitude. The colors indicate different filters and binnings. A fit with an exponential function was performed and the best coefficients are given in the legend and plotted as lines.}
\label{fig:flatfield}
\end{figure}
To perform this task in a fully automatic way, it is best to first measure optimal exposure times as a function of solar altitude. For taking flat-fields, we always point the telescope at the same sweet spot on the sky, right opposite the sun at an altitude of $80^{\circ}$ \citep[see][]{1996PASP..108..944C}. That way, we get comparable count rates for a given solar altitude and exposure time. We take a series of flat-fields, for which we try to get a constant flux level -- in our case $30,000\,\mathrm{counts}$ --, and calculate the optimal exposure time that would be required to get exactly the given level. Figure~\ref{fig:flatfield} shows this for a set of RGBC filters and three different binnings as measured at the 50cm Cassegrain telescope based at the Institute for Astrophysics and Geophysics in G\"ottingen.

As one can see, the measured points do not overlap perfectly over several nights, which can be caused, e.g., by clouds. But the data is good enough to fit exponential functions to it (see lines in plot), which we can use to roughly estimate the optimal exposure time for a given solar altitude.

During dusk twilight, a flat-field module picks a filter and binning combination and estimates the exposure time $t$ for the current solar altitude. If the time is shorter than a given minimum $T_\mathrm{min}$, it does nothing and waits. When $t$ reaches $0.5 \cdot T_\mathrm{min}$, test exposures are started, actually measuring the counts in the image, and calculate a new best exposure time. Only when $t \geq T_\mathrm{min}$ the module starts taking actual flat-fields until either a given number of images has been taken or $t$ gets larger than a given maximum $T_\mathrm{max}$. In dawn twilight, the procedure can be performed accordingly with $T_\mathrm{min}$ and $T_\mathrm{max}$ swapped and opposite comparisons. This is implemented in the \texttt{FlatField} module in \texttt{modules.flatfield}.

The class handling the actual flat-fielding is, again, fully configurable. An example for the \texttt{flat\_fielder} parameter of the module might look like this, defining functions for the exposure time for different binnings and filters:
\begin{verbatim}
    class: pyobs.utils.skyflats.FlatFielder
    pointing:
      class: pyobs.utils.skyflats.pointing.SkyFlatsStaticPointing
    combine_binnings: False
    functions:
      1x1:
        Clear: exp(-1.22421*(h+4.06676))
      2x2:
        Clear: exp(-0.99118*(h+4.66784))
      3x3:
        Clear: exp(-1.14748*(h+5.38661))
\end{verbatim}

The given class for \texttt{pointing} can also be used in the \texttt{FlatFieldPointing} module, which only points the telescope to a specific position without taking flat-fields. This can be useful, if multiple instruments are supposed to be flat-fielded at the same time.

A twilight is usually long enough for taking flat-fields in more than one filter/binning combination. The \texttt{FlatFieldScheduler} module provides a way to run multiple ones as long as the twilight lasts. It can also read priorities from a customizable source, which can be, e.g., an image archive, so that the priorities are the larger the longer ago the last flat-fields in this combination were taken.

\subsubsection{Acquisition}
\label{sec:acquisition}
After moving a telescope to a target, it is often off by some arcseconds or even arcminutes. Sometimes this is unacceptable, especially when the light of a star, e.g., needs to be coupled into a small fiber. In those cases, a fine acquisition based on images from some camera is required. The \texttt{Acquisition} module (in \texttt{modules.pointing}) takes images, runs them through a pipeline (see Section~\ref{sec:images}) to determine what offset to move the telescope, and then applies this offset. This is repeated until the offset is smaller than a given limit.

The configuration for the pipeline typically consists of three steps:
\begin{verbatim}
    pipeline:
      - class: pyobs.images.processors.detection.SepSourceDetection
      - class: pyobs.images.processors.astrometry.AstrometryDotNet
        url: https://astrometry.example.com/
        radius: 5
      - class: pyobs.images.processors.offsets.AstrometryOffsets
\end{verbatim}
First, a source detection is run on the images, followed by an attempt to plate-solve it using the service of Astrometry.net \citep{lang2010}, for which we provide a self-hosted solution (see Section~\ref{sec:pyobs-astrometry}). In the last step, the found coordinates are compared to those from the pointing, and an offset is calculated. Alternative methods are possible by simply changing the pipeline. For instance, an image processor could find the brightest star in the image and set the offset to move the telescope there.

Applying the offset to the telescope is also fully configurable. For a telescope that accepts RA/Dec offsets, it might look like this:
\begin{verbatim}
    apply:
      class: pyobs.utils.offsets.ApplyRaDecOffsets
      max_offset: 3600
\end{verbatim}
The given class simply takes the offsets from the image (written by an image processor) and moves the telescope accordingly.

\subsubsection{Auto-guiding}
The task of auto-guiding is quite similar to that of acquisition, so the configuration is as well: it also mainly consists of a \texttt{pipeline} and an \texttt{apply} step. But instead of running until the calculated offset is small enough, the auto-guiding runs forever, or until stopped, to correct for any shift in pointing that the telescope is doing over time.

In pyobs, two kinds of auto-guiding are ready to use:
\begin{itemize}
    \item \textbf{Science-frame auto-guiding} (module \texttt{ScienceFrameAutoGuiding}) uses the images of the science camera for guiding. This works quite well if the exposure time is small enough to correct for any shifts of the telescope over time.
    \item In contrast, what we just call \textbf{auto-guiding} (module \texttt{AutoGuiding}) requires an extra camera that is mounted, e.g., at the same focal plane as the science camera or at an extra guiding telescope that moves along with the main telescope. In this case, with a bright enough star in the field, the auto-guiding can perform its corrections, independently of the actual science taken, in intervals as short as required.
\end{itemize}

While the astrometric method used in the acquisition would also work for auto-guiding, it is usually too slow. These alternative methods are provided with pyobs:
\begin{itemize}
    \item A projection method as implemented by \texttt{ProjectedOffsets} projects the images separately along x and y axis and cross-correlates both individually with a reference image. The resulting x/y pixel offset can be translated into a RA/Dec or Alt/Az offset.
    \item Cross-correlating full images is usually too slow, so \texttt{NStarOffsets} uses star positions from a source detection that needs to run before, and cross-correlates only small images around the $N$ brightest stars in the image.
\end{itemize}

\subsection{Utilities}
A couple of smaller utility modules for common tasks are provided for convenience.

\subsubsection{Weather}
\label{ssec:utilities.weather}
For fully autonomous observatories, the most important task is not to get observations done, but to close the roof on bad weather and to keep it closed -- an expensive telescope and camera is worth nothing if regularly rained on. With pyobs-weather (see Section~\ref{sec:pyobs-weather}) there is an affiliated project that acts as an aggregator for data from several weather stations and evaluates some logic to determine, whether the weather is good or bad, i.e.\ suitable for observations or not.

The \texttt{Weather} module connects to an instance of pyobs-weather and can provide several functions within a pyobs network:
\begin{itemize}
    \item It provides FITS header entries with weather information for science data.
    \item It has a simple \texttt{is\_weather\_good()} method returning a boolean, indicating whether the weather is good or not.
    \item It sends events when the weather status changes, \texttt{GoodWeatherEvent} and \texttt{BadWeatherEvent}, which other modules can handle and react accordingly.
\end{itemize}

Note that the safety net cast by this module is mainly for the robotic system to react on changes. It is not a replacement for an emergency shutdown in case of, e.g., rain, which should work even without network.

\subsubsection{Telegram}
Even the best logging is only good, if someone reads it. Therefore, the module \texttt{Telegram} can forward all messages of a given level (info, warning, error, ...) into a Telegram chat -- the default configuration would be to have only error messages sent. That way the telescope administrator usually gets notified of a problem within seconds.

The Telegram bot used for this provides several commands that can be issued to it by simply opening a chat on the smart phone. For security reasons, every user has to login (\texttt{/login} command) before receiving any logs and before being able to issue any other command. \texttt{/loglevel} changes the current log level and \texttt{/modules} lists all online modules. The most powerful command is \texttt{/exec}, which allows the user to issue any pyobs command to any module, similar to what is possible within a module or in the Shell of the GUI (see Section~\ref{sec:gui}). Using this, the administrator can easily shut the roof or abort an observation from within a Telegram chat.

\subsubsection{Trigger}
Events are a powerful system in pyobs and for some of them a default action should be performed every time they are encountered. Instead of writing a new module for this, one can simply use the existing \texttt{Trigger} module. It defines events and the method on a given module that should be executed, when the event is triggered. For example:
\begin{verbatim}
    triggers:
      - event: pyobs.events.GoodWeatherEvent
        module: dome
        method: init
      - event: pyobs.events.RoofOpenedEvent
        module: telescope
        method: init
\end{verbatim}
This configuration calls \texttt{dome.init()} on a \texttt{GoodWeatherEvent} and \texttt{telescope.init()} on a \texttt{RoofOpenedEvent}, thus opening roof and telescope when the weather changes from bad to good -- which, in case of pyobs-weather, is usually also the case after sunset for a night telescope. Note that there is no trigger configuration for the bad weather case, since all modules handle that on their own.

\subsubsection{FileCache}
\label{sec:filecache}
While a camera module can be configured to store its files locally, that can be quite impractical, if it runs on a different computer than the rest of the pyobs system, which might be the case quite often. So there is need for a place to store the images that can be accessed from all modules -- or at least those that need access to the images.

A network mount using, e.g., SMB or NFS does the job well, but with \texttt{HttpFileCache} there is also a module available for that in pyobs. It opens a web server on a given port, which can be used to upload images from the camera and download them somewhere else. It can simply be accessed via the VFS (see Section~\ref{sec:vfs}) using a \texttt{HttpFile} root.

\subsubsection{ImageWriter and ImageWatcher}
When the camera uploads its images to a FileCache (see above), they should still be stored somewhere, since the cache only holds a limited amount of files. An easy way to do that is the \texttt{ImageWriter} module that waits for \texttt{NewImageEvent}s, downloads those images and stores them at a different VFS location.

To make this a little safer and reduce the risk of losing images, an \texttt{ImageWriter} should always write images to a local disk. If they are supposed to be copied to a remote location, the preferred way is an additional \texttt{ImageWatcher}, which watches a given path for new files, copies the files somewhere else, and only deletes the original files if there was no error. So a typical setup would configure the \texttt{ImageWriter} to store its files into a local directory like this, assuming that the camera stores its images at \texttt{/cache/} and \texttt{/some/temp/dir/} is some local temp directory:
\begin{verbatim}
    class: pyobs.modules.image.ImageWriter
    vfs:
      class: pyobs.vfs.VirtualFileSystem
      roots:
        cache:
          class: pyobs.vfs.HttpFile
          download: http://somewhere:37075/
        archive:
          class: pyobs.vfs.LocalFile
          root: /some/temp/dir/
\end{verbatim}
Note that the root \texttt{archive} is used since the default value for the \texttt{filenames} parameter of the module is \texttt{/archive/{FNAME}}.

After the images have been stored locally, an \texttt{ImageWatcher} should pick them up and copy them into an archive (note that curly brackets in \texttt{destinations} indicate placeholders which are filled from FITS header values):
\begin{verbatim}
    class: pyobs_iagvt.filewatcher.FileWatcher
    watchpath: /temp/
    destinations:
      - /archive/{FNAME}
    vfs:
      class: pyobs.vfs.VirtualFileSystem
      roots:
        temp:
          class: pyobs.vfs.LocalFile
          root: /some/temp/dir/
        archive:
          class: pyobs.vfs.ArchiveFile
          url: https://archive.example.com/
\end{verbatim}
Only after the images have been copied into the archive they will be deleted from the temp directory.

\subsection{GUI}
\label{sec:gui}
While all the other modules presented here are fully autonomous, pyobs also provides a graphical user interface (GUI) for easy (remote) access to the system. Technically it is also just another module, which opens a window for interaction with the user.

\begin{figure}[t]
  \begin{center}
    \includegraphics[width=\textwidth]{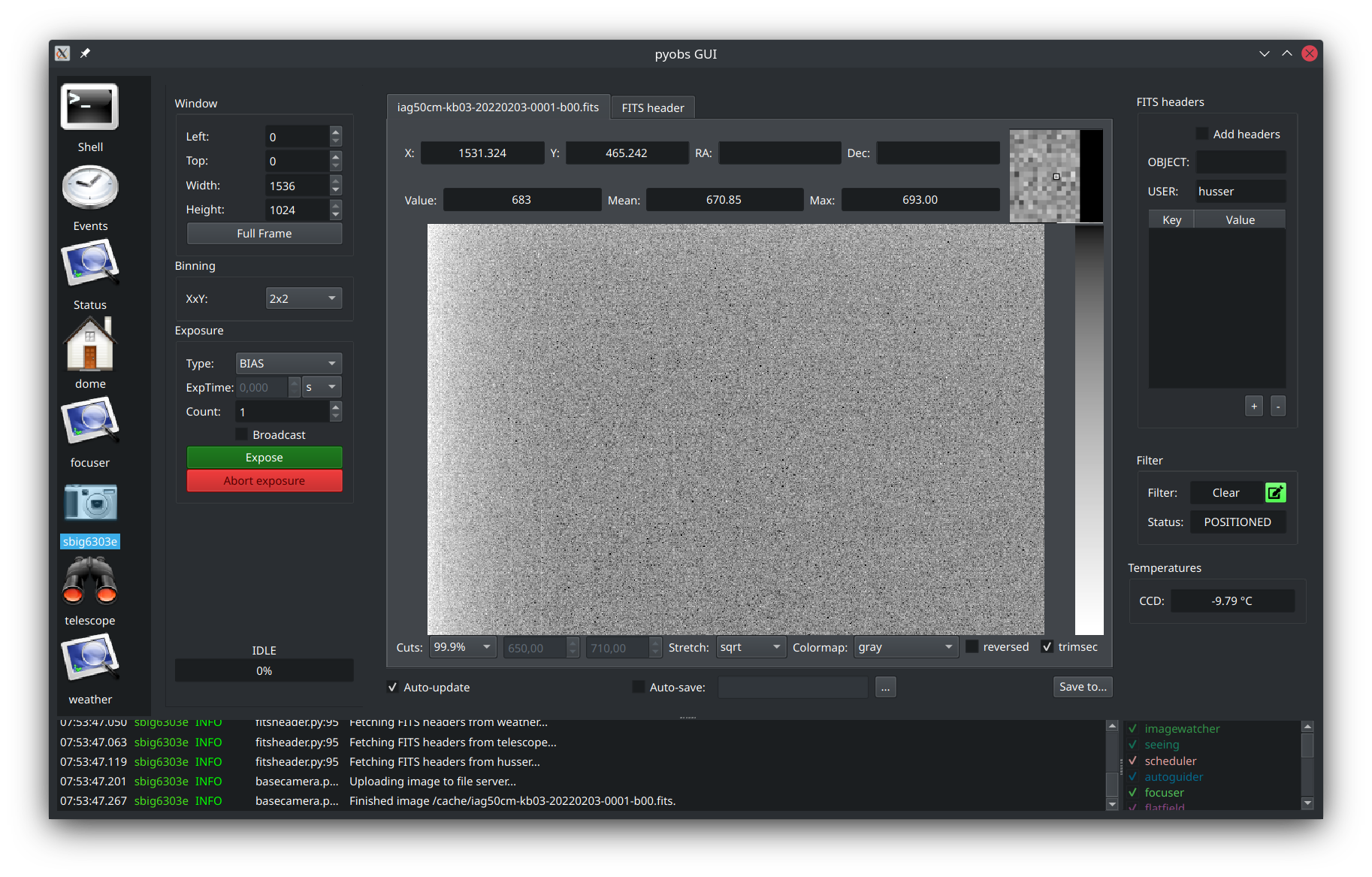}
  \end{center}
  \caption{A screenshot of the graphical user interface (GUI) as provided by pyobs-gui. It shows the list of connected modules that are supported by the GUI on the left. When selecting one, a custom widget for each kind of module is shown in the main area right of it. Below is the logging area, which shows log entries from all connected modules.}
  \label{fig:gui}
\end{figure}
Figure~\ref{fig:gui} shows a screenshot of the GUI right after a bias image has been taken with the selected SBIG camera. The main window of the GUI consists of three major parts:

\begin{itemize}
    \item The \textbf{list of module pages} on the left, including the three special pages Shell, Events, and Status.
    \item The \textbf{system log} on the bottom, showing all log entries from all connected modules as well as a list of all those modules on the lower right.
    \item The \textbf{module page}, filling the rest of the window, which changes depending on the selected module.
\end{itemize}

The three special pages mentioned above are:

\begin{itemize}
    \item The \textbf{Shell} is an interactive command prompt, in which the user can execute any command on any module in the form \texttt{<module>.<method>(<params>)}. This makes the shell a very powerful tool for admins and for debugging.
    \item The \textbf{Events} page shows a chronological list of all events that have been sent in the pyobs network. It also allows to send events on its own with parameters defined by the user.
    \item The \textbf{Status} page shows the current status of a module, e.g., whether it is in an error state. It also shows the pyobs version of every module to keep track of updates.
\end{itemize}

In the list of modules on the left, not all modules are listed, but only those for which a graphical user interface has been designed. The GUI is fully dynamic, which means that it changes according to the list of connected modules. Single module pages also adapt to the capabilities of the associated module, e.g.\ the camera page only shows options for window and binning, if the camera supports it.

The customization of the GUI goes even further with user-defined pages. For example, pyobs does not provide a user interface for acquisition and guiding, but in the case of our solar telescope, a visual feedback is important. So we created a new widget and defined it in the configuration of the GUI:
\begin{verbatim}
    widgets:
    - module: guiding
      overwrite: True
      widget:
        class: pyobs_iagvt.guidingwidget.GuidingWidget
        acquisition: acquisition
\end{verbatim}
This tells the GUI to overwrite an existing widget for the \texttt{guiding} module with the given class. Using custom widgets, one can adapt the GUI to work with any special requirements.

The other way around, restricting access in the GUI, can also be accomplished in the configuration via the \texttt{show\_shell}, \texttt{show\_events}, and \texttt{show\_status} parameters, which, if set to False, hide the corresponding page. An explicit list of allowed module pages can be provided with the \texttt{show\_modules} parameter. Here is an example for a very limited access to the camera only:
\begin{verbatim}
    show_shell: False
    show_events: False
    show_status: False
    show_modules: [camera]
\end{verbatim}
Altogether, the GUI tries to allow access to all modules as well as it can, but it is also highly customizable to match any requirements of an observatory. With ports for the XMPP server (and probably the file cache) open to the public, this enables a safe and easy remote access to the pyobs system.

\section{Affiliated projects}
There are a few projects with "pyobs" in their name that do not provide any new modules but some external services that are essential for operating a fully-autonomous telescope.

\subsection{Image archive}
\label{sec:pyobs-archive}
In classic astronomy an observation consists of three steps:

\begin{enumerate}
    \item Planning an observation, i.e.\ finding targets, defining filters and exposure times, evaluating best times for the observation, etc.
    \item Actually performing the observation at the telescope.
    \item Calibrating and analyzing the data.
\end{enumerate}

Nowadays it is absolutely possible to automate all three and avoid human interaction at all. While this topic goes far beyond the scope of this paper, we want to mention that a full automation does not only work for large surveys, but also for small telescopes in the middle of a town like G\"ottingen (see Masur et al., in prep). However, for robotic observations at least the second step falls away from the observer's responsibility, but also parts of step one (defining observing times) and three (calibrate data). In that case, probably not knowing exactly when an observation was taken, an efficient way to find data becomes more important.

This is where an image archive comes into play. There is the LCO science archive\footnote{\url{https://archive.lco.global/}}\footnote{\url{https://github.com/observatorycontrolsystem/science-archive}} to use, but it stores the images in Amazon AWS S3, while we wanted to store data locally. So we developed our own backend, which also supports the LCO API, and took parts of the LCO web frontend with permission and adapted it to our needs. We also added a HTTP endpoint for uploading images. Within pyobs there are classes for both an easy upload using the VFS (via \texttt{ArchiveFile}), and a full wrapper for accessing the archive in \texttt{PyobsArchive}.

\begin{figure}[t]
\begin{center}
\includegraphics[width=\textwidth]{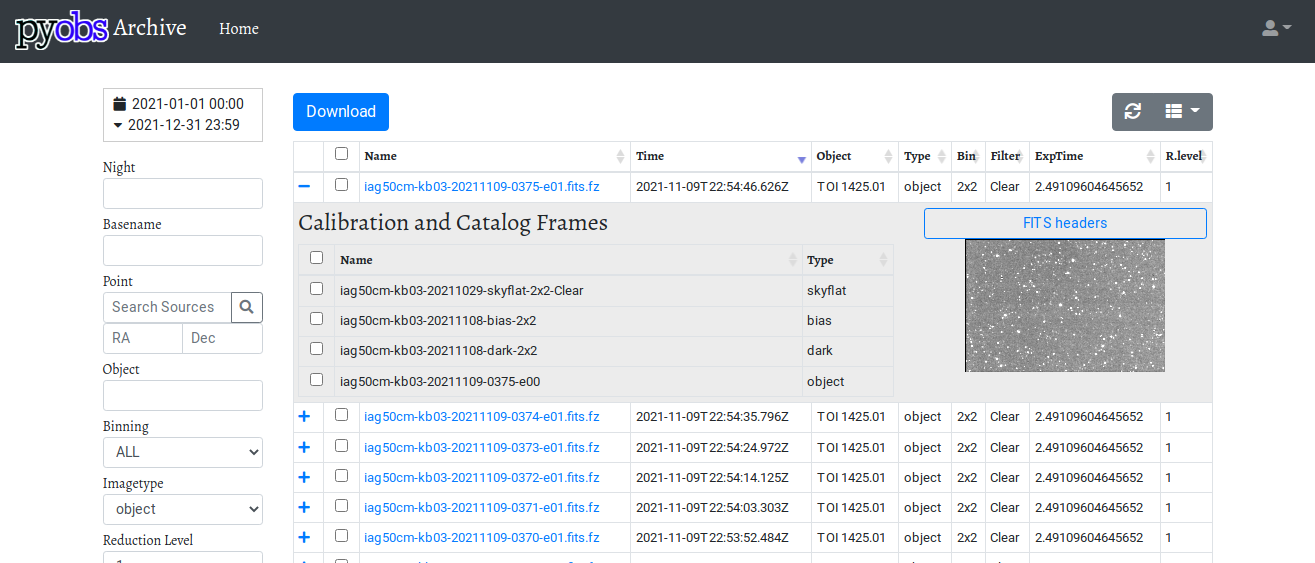}
\end{center}
\caption{Screenshot of the pyobs-archive instance that we use for the MONET telescopes and the IAG50cm telescope.}
\label{fig:pyobs-archive}
\end{figure}
Figure~\ref{fig:pyobs-archive} shows a screenshot of the archive that we use for the two MONET telescopes and the IAG50cm telescope. On the left, there is a list of options to filter the data by. On the right is the list of images matching the selected criteria. More details -- including connected data (for calibrated images), a link to the FITS headers and a thumbnail preview -- can be accessed by clicking on the plus symbol. Single or multiple images can also easily be downloaded using this web frontend.

\subsection{Weather aggregator}
\begin{figure}[t]
\begin{center}
\includegraphics[width=0.49\textwidth]{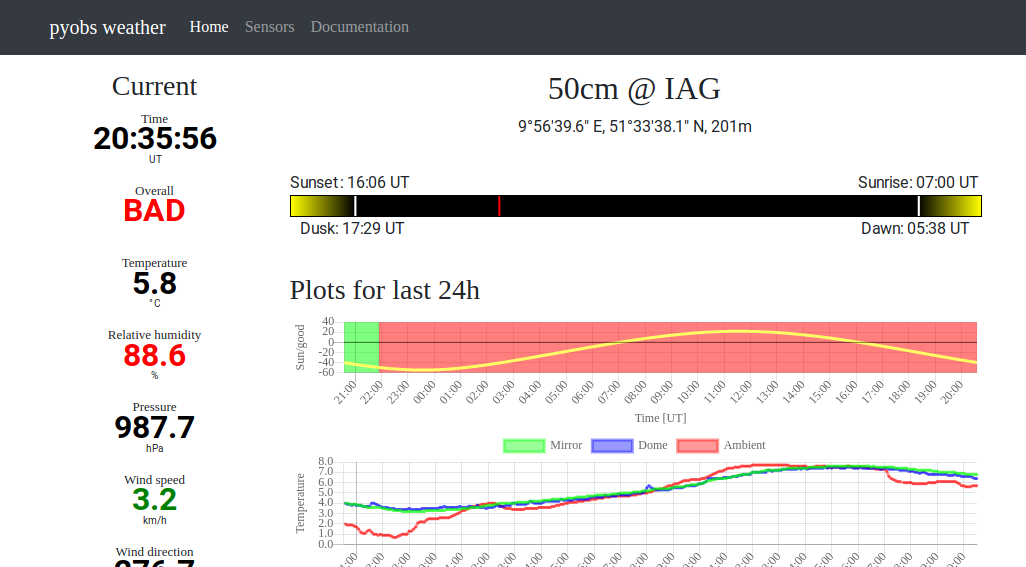}
\includegraphics[width=0.49\textwidth]{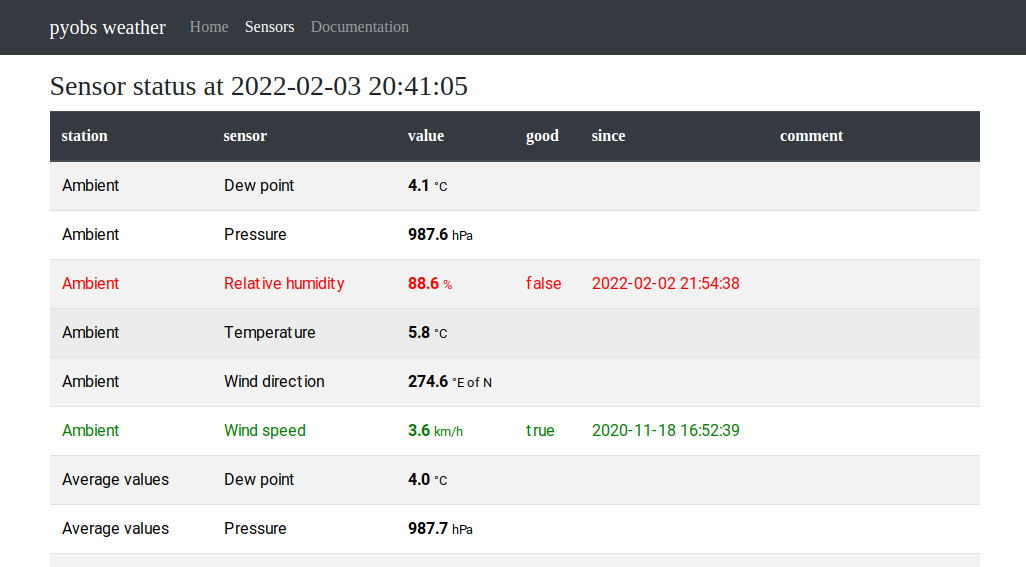}
\end{center}
\caption{Two screenshots from the pyobs-weather web frontend for the IAG 50cm telescope. The main page is shown on the left and the Sensors page on the right.}
\label{fig:pyobs-weather}
\end{figure}

In Section~\ref{ssec:utilities.weather} we already mentioned a project for aggregating data from different weather stations and evaluate the values in order to determine, whether the weather is good for observing. Figure~\ref{fig:pyobs-weather} shows two screenshots from pyobs-weather as used by the IAG50cm telescope.

On the left, the main page is shown, with average values from all sensor types as well as plots for the current night (top), weather status (green and red shaded areas in the plot below, indicating good or bad) and solar altitude (yellow line in same plot), and plots for all sensors, grouped by type (i.e..\ temperature, humidity, etc). On the right, the Sensors page is shown with current values for all sensors from all stations, times of last changes (for evaluated sensors, see below) and comments on the current status. There is a public API for the weather data, which can easily be accessed via the weather module (see Section~\ref{ssec:utilities.weather}).

The system is fully customizable. The basic unit in pyobs-weather is a \textit{station}, which usually defines a single physical weather station. There are some station classes already present, of which some are more generic (getting data from a MySQL or CSV table) and some are specific to the observatories where our telescopes are located (e.g.\ for the weather station on Mt. Locke at McDonald Observatory). Additional classes for more stations can easily be added. There are three special stations that do not represent an actual weather station: \texttt{Current} contains the current average values from all stations, while \texttt{Average} keeps a five-minute average. Finally, \texttt{Observer} calculates current conditions, which, at the moment, is only the solar altitude.

Every station then contains one or more \textit{sensors}, which provide values for a sensor of a given type: temperature, relative humidity, pressure, wind speed and direction, particle count, rain, sky temperatures, or solar altitude. To each sensor, one or more \textit{evaluators} can be attached, which take the current value and decide whether it allows for observations or not. Currently, pyobs-weather offers four different kinds of evaluators:
\label{sec:pyobs-weather}
\begin{itemize}
    \item A \texttt{Boolean} is a simple logic evaluator, which is \texttt{True} if the sensor value is \texttt{True}, and vice versa -- or the opposite, if \texttt{invert} is set to \texttt{True}.
    \item \texttt{Switch} is a simple switch, which is \texttt{True}, if the sensor value is above a given \texttt{threshold}, and vice versa. And, again, the other way around, if \texttt{invert} is set to \texttt{True}.
    \item A \texttt{Schmitt} trigger is similar to a \texttt{Switch}, but it takes two values: for it to become \texttt{True}, the sensor value must be below a given \texttt{good} value, but to become \texttt{False} again, it must rise above a given \texttt{bad} value.
    \item Sensor values have a \texttt{valid} flag, which is mostly used (and set to \texttt{False}), if the value is older than five minutes. The \texttt{Valid} evaluator only evaluates to \texttt{True}, if the value is valid.
\end{itemize}

As an example, we assume getting relative humidities from two weather stations. For both we would typically set a \texttt{Schmitt} evaluator with values like \texttt{good}=80 and \texttt{bad}=85, which means that the weather is marked as bad, if the humidity rises above 85\%, but is only marked good again, if the humidity falls below 80\%. It would not be a good idea to attach a \texttt{Valid} evaluator to both, since weather stations can break. However, we still always want a valid reading for the humidity, so we assign it to the humidity sensor in the \texttt{Current} station. That way, if we get no valid value at all, the weather is marked as bad. Evaluators on the \texttt{Average} station are never evaluated, but they are used for color coding the plots, i.e.\ mark areas that would mean bad weather.

\subsection{Astrometry}
\label{sec:pyobs-astrometry}
Getting astrometric solutions for images (i.e.\ "plate-solving" them) is a task required at multiple occasions (see, e.g., Sections~\ref{sec:images}~\&~\ref{sec:acquisition}). For this we use a self-hosted version of Astrometry.net \citep{lang2010}, adding a HTTP interface for accessing its service. Similar to Astrometry.net's own web service,\footnote{\url{http://nova.astrometry.net}} it accepts a list of X/Y positions of stars on an image, but in addition some parameters for the fit can be provided, like a first guess for the coordinates and an estimate for the plate-scale. A successful call returns FITS header entries that can be added to an existing file in order to get a valid world coordinate system (WCS). The whole process usually takes well below one second. pyobs provides an image processor that uses this web service for easy use in a pipeline (see Section~\ref{sec:images}).

\section{Full robotic mode}
With everything described so far, we already have a working observatory. We can control all devices, automate some things, and remotely control the system with the GUI. All that is needed for a fully autonomous telescope is some piece of software that coordinates everything. These robotic systems come in all shapes and colors: from a rather simple survey mode, in which a pre-defined list of targets is executed from top to bottom, probably all to be done with the same settings, whenever the conditions are right, to a system with user-defined tasks, maybe multiple instruments, and a scheduler that tries to fit all together.

\subsection{Scheduling}
\label{sec:scheduling}
The most simple robotic system imaginable is a simple list of targets that are to be observed one after the other, top to bottom. An algorithm for that might look like this:
\begin{enumerate}
    \item Select a target from a list, probably the first one.
    \item Move the telescope to the given coordinates.
    \item Take an image and store it.
    \item Repeat.
\end{enumerate}

A system like this still has some other things to take care of, e.g.\ open up at dusk (for night observations) and close down at dawn -- or when the weather gets bad. Any interruption (like daylight or rain) would just delay the selection of the next target. While very simple, this kind of system is suitable for many types of observations. There is no module implementing a survey mode in pyobs, but it could easily be added with very few lines of new code, specialized on the use case at hand.

This "survey mode" is also easily extendable, e.g., add an exposure time and a filter to the table of targets and set them before starting the exposure. However, the targets would still be observed in the order that they appear in the table. Therefore, the next step might be to filter the table of targets by visibility and sort it by some kind of priority. If we do that every time the system is idle, we get some kind of "just-in-time" (JIT) scheduler, always picking the next target when needed, but never planning further ahead. Some control systems, like the one for STELLA on Tenerife \citep{2012ASInC...7..247G}, developed this idea further and have been using it successfully for years. A JIT scheduler can be very powerful, because it can easily adapt to changing observing conditions like seeing or transparency and picks its next target accordingly.

There is, however, one major disadvantage for these kind of systems: selecting only the next targets means there is no full plan for the night (or day), so it may be difficult to impossible to predict, whether a specific target will be observed or not. It may even be difficult to decide, which parameters need to be changed in order to make sure the observation will take place. Furthermore, the selection of targets may never be "optimal", i.e.\ it is challenging to fill the observing time with the best possible targets. For example, take an object $A$ that can be observed at the beginning and at the end of the night (maybe a transit event). Another object $B$ can only be observed at the beginning of the night. Even if $A$ has a higher priority, it might be better to observe $B$ first and then $A$ at the end of the night.

An astronomer, going on an observing run, would probably plan the nights in advance and make a schedule, when to observe which target. This is an optimizing problem and so we can call these kinds of schedules "optimal". This is a different approach to selecting targets and not as straightforward as the one for JIT schedulers described before. Luckily, there are free schedulers available for use, e.g.\ the adaptive scheduler developed by LCO\footnote{\url{https://github.com/observatorycontrolsystem/adaptive_scheduler}} and the Astropy-affiliated project astroplan, just to name two. In principle, they all try to optimize the placement of observations for the whole night so that a given value is maximized, e.g.\ the total observing time or something like the time-integrated priorities of the tasks.

While the LCO scheduler can run fully independent from pyobs, there is a module based on astroplan: \texttt{Scheduler}. It takes schedulable tasks from a \texttt{TaskArchive} object, calculates a schedule, and writes it to a \texttt{TaskSchedule} object. A \texttt{TaskArchive} simply holds a list of tasks (in the form of \texttt{Task} objects) and returns them on request. The scheduler takes these tasks, converts them into astroplan's \texttt{ObservingBlock}s, applies given constraints, and starts the scheduler. The result is a time table, giving start and end times for all scheduled tasks, which is passed to the \texttt{TaskSchedule}, storing it to be accessed by the robotic telescope system.

All these three classes (\texttt{TaskArchive}, \texttt{TaskSchedule}, and \texttt{Task}) are abstract and need specific implementations for a method to store tasks and schedule. The implementation coming with pyobs is one tailored to be used with the LCO observing portal, but access to other task databases can easily be added.

Furthermore, this gives a simple framework for changing the used scheduler at a later time. The one implemented in astroplan is a "greedy" one, i.e.\ it schedules the task with the highest priority first, then the one with the next lower priority, and so on. While it ensures that the highest priority target is observed, this is not true for all other targets. Thus, the result of a "greedy" scheduler is still far away from an optimal one. However, changing the actual scheduler will not affect the rest of the robotic system at all.

\subsection{LCO observing portal}
The central part of the LCO observing portal is a database, mainly storing tasks, schedules, and observations, and a HTTP REST interface for accessing it -- see details about the API on LCO's developers page\footnote{\url{https://developers.lco.global}}.

When the portal is set up correctly and running, a new account must be created with "Staff" permissions to access all the necessary endpoints. The security token for this account must be provided in the configuration of \texttt{LcoTaskArchive} and \texttt{LcoTaskSchedule}, which are the LCO-specific implementations of the classes discussed above. They both make use of \texttt{LcoTask} that simply stores the JSON object returned from the portal. These classes are enough to run the scheduler in connection with an LCO portal.

\begin{figure}[t]
  \begin{center}
    \includegraphics[width=\textwidth]{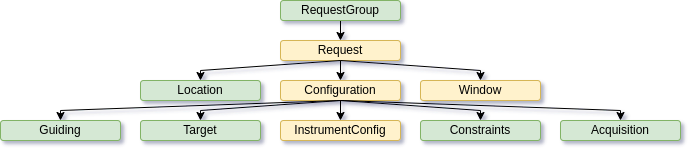}
  \end{center}
  \caption{Structure of a task in the LCO portal. While the green fields can occur only once, there can be multiple entries for the yellow fields (Request, Configuration, Window, InstrumentConfig).}
  \label{fig:flowchart_lco}
\end{figure}

Figure~\ref{fig:flowchart_lco} shows the structure of a task in the LCO portal:
\begin{itemize}
    \item The top-most element is a request group, which has a name and belongs to a proposal. It can contain one or more requests.
    \item A request contains a location (i.e.\ the telescope to use), one or more observation windows and one or more configurations.
    \item A configuration stores settings for acquisition and guiding, observing constraints (airmass, moon distance, etc.), the target information and one or more instrument configurations.
    \item Finally, an instrument configuration holds information like exposure time and count, and filter to use, all depending on the selected instrument.
\end{itemize}
Each of these elements also contains an \texttt{extra\_params} field, which can be used for any extra information that is not supported by the default parameters.

Configurations have a \texttt{type} parameter, which will be important for running the task. The default value for an imaging camera would usually be \texttt{EXPOSE}, which just exposes as many images as given in the instrument configuration. Another possibility is \texttt{REPEAT\_EXPOSE}, which loops all instrument configurations, until a given \texttt{repeat\_duration} is reached. There are also other, more specific types, like \texttt{AUTO\_FOCUS} for performing an auto-focus series.

\subsection{Running tasks}
With the schedule in place, we actually need to observe the tasks. In pyobs this is done by a \texttt{TaskRunner}, which only has two methods: \texttt{can\_run()} checks, whether a given task can run right now, and \texttt{run\_task()} actually executes it. For this, pyobs uses the concept of "scripts", which can be fully customized in the \texttt{runner} section of a configuration for a task runner. While the following will concentrate on running tasks from an LCO portal, implementations for other task archives should be easily implemented.

In the case of the LCO portal, the script to use is defined by the configuration type. A possible configuration might look like this:
\begin{verbatim}
    runner:
      class: pyobs.robotic.TaskRunner
      scripts:
        BIAS:
          class: pyobs.robotic.lco.scripts.LcoDefaultScript
          camera: sbig6303e
        EXPOSE:
          class: pyobs.robotic.lco.scripts.LcoDefaultScript
          telescope: telescope
          filters: sbig6303e
          camera: sbig6303e
          roof: dome
          acquisition: acquisition
          autoguider: autoguider
\end{verbatim}
This uses the same script (\texttt{LcoDefaultScript} in \texttt{robotic.lco.scripts}) for two different configuration types. The script checks internally, what kind of observation to perform. As usual, all parameters in the configuration are forwarded to the constructor of the given class -- in this case all of them are names of modules handling specific tasks. Note that a request can contain multiple configurations, and each might use a different script and might be observable or not.

An \texttt{LcoTask} checks and runs all configurations in a request. The procedure that the \texttt{LcoDefaultScript} runs for a single configuration looks like this:
\begin{itemize}
    \item If the configuration type is \texttt{EXPOSE}, which is a science observation, move the telescope to the given coordinates.
    \item If a fine acquisition is requested, do it.
    \item If guiding is requested, start it.
    \item Now loop all instrument configurations and for each set the filter and binning and take the given number of images.
\end{itemize}
If the configuration type is \texttt{BIAS} or \texttt{DARK}, this procedure would simplify to just taking images. 

In the case of this \texttt{LcoDefaultScript}, the script needs to know about internals of the task that are not available via the public interface of \texttt{Task}, so an LCO specific script is required. The same is true for the auto-focus script \texttt{LcoAutoFocusScript}. On the other hand, there are some scripts that do not need any information from the task and can therefore be run in any system, e.g.\ the \texttt{SkyFlats} script. When using this in a LCO environment, the configuration type \texttt{SCRIPT} is required and a special script \texttt{LcoScript}, which evaluates a \texttt{script} parameter in the \texttt{extra\_params} of the configuration to delegate execution to another script. It can be configured like this:
\begin{verbatim}
    runner:
      class: pyobs.robotic.TaskRunner
      scripts:
        SCRIPT:
          class: pyobs.robotic.lco.scripts.LcoScript
          scripts:
            skyflats:
              class: pyobs.robotic.scripts.SkyFlats
              [...]
\end{verbatim}
Now, whenever the configuration type is \texttt{SCRIPT} and the \texttt{script} is set to \texttt{skyflats}, the given script \texttt{SkyFlats} is executed. The whole script system is designed to be as flexible as possible and should allow for writing custom scripts for any requirement.

For our solar telescope we also use a (modified) LCO portal, but the robotic mode is a lot simpler: there is no scheduler, but the task archive just requests the schedulable blocks and returns the one with the highest priority. This is possible, because all positions on the solar disc are visible as soon as the sun is up in the sky. There is also a different default script that just moves the telescope and triggers the spectrograph.

\subsection{The mastermind}
Using the scripts system, building a central module that runs them becomes very simple -- we call this module the "mastermind". It creates a \texttt{TaskSchedule} and a \texttt{TaskRunner} from its configuration and then continuously gets the tasks from the former and executes them via the latter. It also sends events when starting and finishing a task and writes information about the task into the FITS headers of the images.

The whole system is flexible enough that we run two 1.2m and one 0.5m night telescope with it, as well as a 0.5m solar telescope -- however, for the last one the default scripts are not used (they use, e.g., a different coordinate system) and even the LCO portal had to be adapted for this use case. Nevertheless, the changes were minimal and we can use the same code base for all telescopes.

\section{Telescopes}
\label{sec:telescopes}
The Institute for Astrophysics and Geophysics in G\"ottingen (IAG) operates four telescopes, of which two are located within the faculty building, one is in Texas, and the last one is in South Africa. In this section we will describe the hardware for each one and their level of automation with pyobs.

\subsection{IAG 50cm}
The IAG 50cm is a Cassegrain telescope located on the roof of the institute with a main mirror with $0.5\,\textrm{m}$ diameter and a focus length of $5\,\textrm{m}$ (f/10), housed in a classical rotating dome. The telescope is mainly used for educational purposes and public outreach. With the use of pyobs we now also use the rare days of good weather in G\"ottingen for science observations, but it is mainly a testing platform for the two MONET telescopes (see below). The main instrument is a SBIG STL-6303E with a pixel scale of $0.55\,"/\textrm{px}$ (with a focal reducer). Attached to the the telescope is a smaller 110mm f/7 refracting telescope with a ZWO ASI 071MC camera.

Dome, telescope and focusing unit are running with ASCOM and are connected to pyobs via \texttt{pyobs-alpaca}. The two cameras use their respective modules (\texttt{pyobs-sbig} and \texttt{pyobs-asi}). The other modules that we use perform the following tasks (for all see Section~\ref{sec:modules}):
\begin{itemize}
    \item Fine acquisition with both of the cameras,
    \item auto-focus for the main telescope and camera,
    \item flat-fielding for both cameras,
    \item file cache and image writer and watcher,
    \item scheduler and mastermind for robotic mode,
    \item telegram bot,
    \item weather from a connected \texttt{pyobs-weather} (see Section~\ref{sec:pyobs-weather}) page.
\end{itemize}

The main telescope runs fully robotically with a copy of the LCO portal (shared with the MONET telescopes), while the smaller telescope is not yet supported in this mode. We are currently working on guiding with the small telescope and on implementing the necessary pointing model in pyobs. We are also currently adding a fiber pick-up to transfer the light from the main telescope to a spectrograph in the optical lab (see next section). For this, the guiding uses a camera from The Imaging Source pointed at the fiber pin hole in a $45^{\circ}$ mirror, which has already been tested (see Fig.~\ref{fig:resolvedsun}, right).

\subsection{IAG VVT}
The Vakuum-Vertikalteleskop (VVT) consists of  a siderostat on the top of the faculty building, redirecting the light two stories down into the building, where the $0.5\,\mathrm{m}$ primary mirror is reflecting the light back up one story and into the optical lab. It provides both a f/11 primary focus and a Gregory f/50 secondary output. 

\begin{figure}[t]
  \begin{center}
    \includegraphics[width=0.49\textwidth]{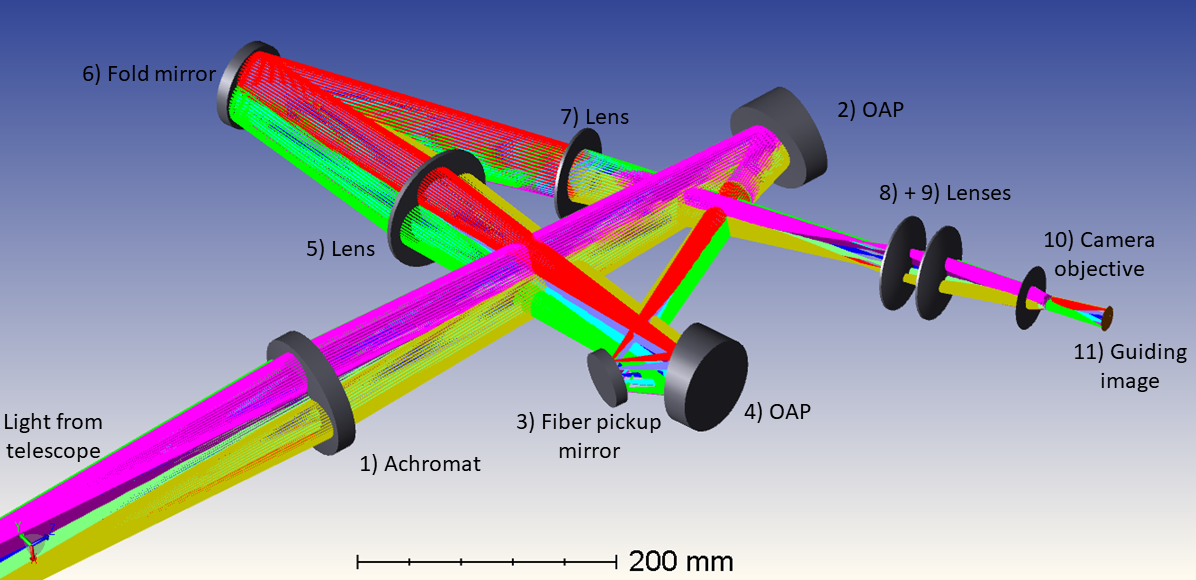}
    \includegraphics[width=0.49\textwidth]{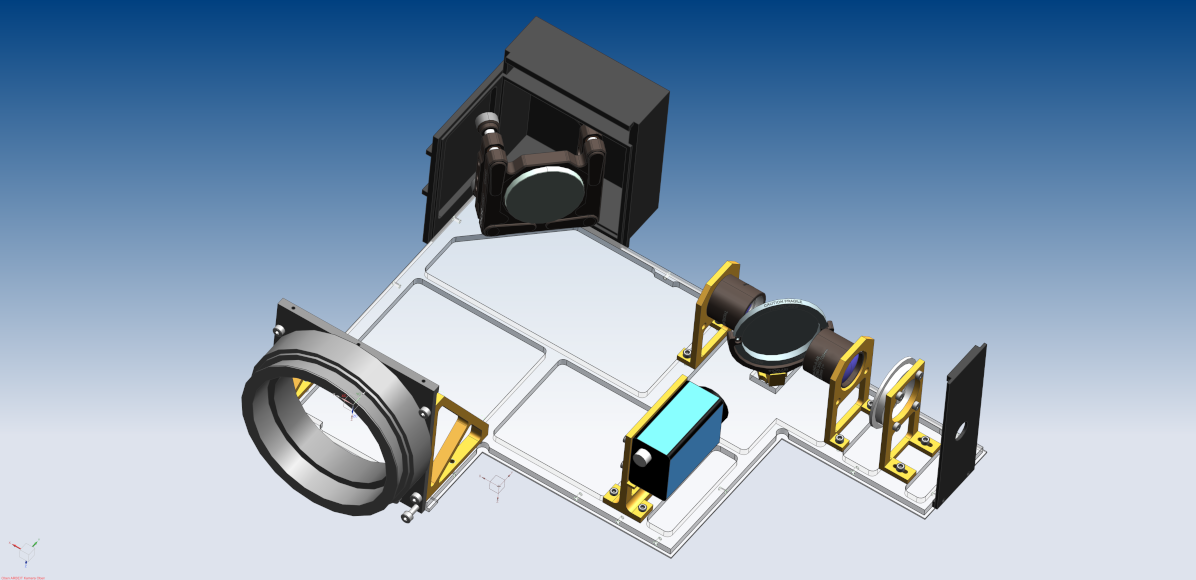}
  \end{center}
  \caption{Left: Mid-resolution resolved Sun fiber setup for the VVT: The full image of the Sun is re-imaged onto the fiber pickup mirror that is hosting a $525\mu\textrm{m}$ fiber (corresponding to a 32 arcsec field of view). The fiber leads to a Fourier-Transform-spectrograph. Behind the pickup mirror the light is again re-imaged, this time onto the guiding camera which is used by pyobs for both pointing and guiding.\\ 
Right: CAD-model of the fiber-guiding unit for the 50\,cm telescope. Starlight is re-imaged onto a fiber-pickup mirror and the remaining light is redirected into the guiding camera, allowing for nearby stars to act as guidestars.}
  \label{fig:resolvedsun}
\end{figure}

There are a total of six observing modes for the telescope, with five of them using pyobs for pointing and guiding using a custom module via an interface to its control system. These modes include different spatial resolved observing modes (with field of views between about 4 and 100 arcsec) and Sun-as-a-star integrated light modes. As an example, Fig.~\ref{fig:resolvedsun} (left) shows a Zemax raytracing of the mid-resolution resolved Sun fiber setup. The light from the primary mirror is collimated and re-imaged onto a fiber pickup mirror and re-imaged a second time onto the CCD guiding camera that is used for acquisition and guiding via detecting the solar disk. 

The light entering the fiber is sent to our Fourier-Transform-spectrograph (FTS), a Bruker IFS 125HR with a maximum resolving power of $>700,000$ at $600\,\textrm{nm}$. For the FTS, another custom module is used for HTTP communication with a LabView instance, which in turn is connected to the instrument software OPUS. For more details on the resolved Sun setup -- see \cite{10.1117/12.2560156}, and for more details on the coupling into the FTS see \cite{10.1117/12.2561599}. We are currently commissioning the fully robotic mode, based on a modified LCO portal, which now accepts coordinates in the Stonyhurst Heliographic system (HGS).

\subsection{MONET/N}
The two MONET Alt/Az telescopes \citep{2001ASPC..246...13H,2006SPIE.6270E..1QB} have (almost) identical hardware with a $1.2\,\textrm{m}$ main mirror at f/7. They were optimized for fast operations with up to $10^{\circ}/\mathrm{s}$ on both axes and therefore also have a clam-shell roof that opens completely. The northern telescope, MONET/N, is located at McDonald Observatory in Western Texas, USA. Its science camera will soon be replaced with a FLI KL4040 and we are in the process of designing a fiber-fed high resolution spectrograph for high-precision radial velocity observations of G-type stars on the m/s-level. 

The level of automation is about the same as for the IAG 50cm, with the exception of the piggyback telescope.

\subsection{MONET/S}
With MONET/S, located at the South African Astronomical Observatory (SAAO) near Sutherland, South Africa, having mostly the same hardware as MONET/N, there are still some differences. The science camera is currently a FLI PL230 and there is a $0.25\,\textrm{m}$ f/8 piggyback telescope mounted at the (unused) second Nasmyth port, with a SBIG STX-8300M camera attached to Gemini derotator. Furthermore, outside the field of view of the main camera we installed a pickup for a fiber bundle leading to MORISOT, a low-budget, low-resolution spectrograph.

\begin{figure}[t]
  \begin{center}
    \includegraphics[width=\textwidth]{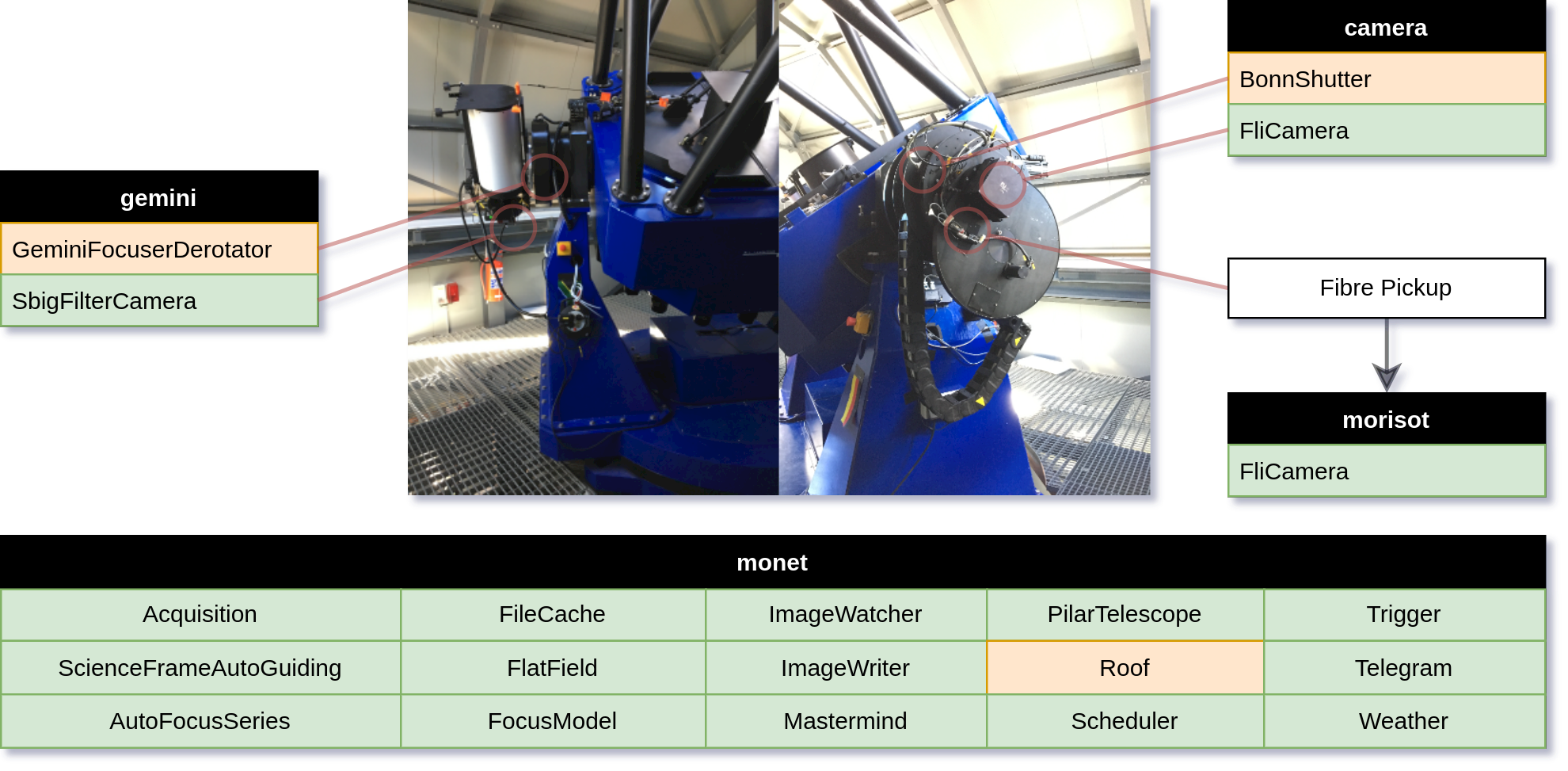}
  \end{center}
  \caption{List of all pyobs modules that are currently running at MONET/S and how they are distributed to four different computers. Marked in green are those modules that are available in pyobs-core or one of the additional packages. The three other modules are custom implementations for the given hardware.}
  \label{fig:monets}
\end{figure}

Again, the level of automation is similar to its twin in Texas and the IAG 50cm. Figure~\ref{fig:monets} shows an illustration of all pyobs module running at MONET/S, how they are distributed over several computers, and how they are connected to the actual hardware. As one can see, there is an additional module for the derotator of the piggyback telescope that we will publish as soon as it is fully tested. Acquisition (with an offset) and guiding of the spectrograph is supposed to be done with the science camera, and we hope to be able to do parallel photometry of the target with the piggyback. The custom module for the roof simply calls HTTP REST endpoints on our roof controller, and \texttt{BonnShutter} continuously checks the health of our Bonn shutter, and resets it, if any error occurs.

When everything is finished, there will probably be three modules for acquisition (one for each instrument) and three modules for guiding: science-frame auto-guiding on the main camera, guiding via piggyback, and guiding via science camera for the spectrograph. There will also be auto-focusing for both telescopes and flat-fielding for all three instruments. A challenge will be to calibrate all three instruments during twilight, but with the flexible scripts in the robotic part of pyobs, this should not be too much of a problem.

\section{Summary \& outlook}
In this paper we presented the observation control system pyobs. While pyobs itself is written in Python and highly depends on third-party packages, it can easily be extended by any programming language that supports the communication protocol XMPP.  We showed that pyobs is highly customizable due to its configuration files, and provides a lot of functionality for robotic telescope operations out of the box: it has support for common tasks like flat-fielding and auto-focus series and connects to the open-source LCO observation portal for organizing tasks.

At the time of writing this paper, pyobs is available in version 0.17. As the leading zero suggests, we do not believe that it has reached a stable state, in which no major changes to any of its system will happen in the near future. However, at least the currently planned modifications are mostly minor, and we expect to publish a first release this year or soon thereafter -- so this should not keep anyone from using pyobs before that.

The list of planned improvements for the core of pyobs is long, but contains mostly minor items, which probably will not affect running systems. Some of the more major ones are:
\begin{itemize}
    \item The error handling (see Section~\ref{sec:error_handling}) is quite new and not used everywhere. It is missing, especially, in the robotic modules.
    \item Some access control will be added, so that a module can allow some of its methods to be called only by authorized clients.
    \item There already exist a few unit tests for the core package, but they are not covering everything, not even the most important parts.
    \item New interfaces (see Section~\ref{sec:interfaces}) will be added -- e.g.\ for supporting to track non-sidereal targets --, which might make it necessary to change existing ones.
\end{itemize}

While these items are for the core system, future development will mainly concentrate on additional modules. For instance, we would like to guide using a guiding telescope, which would require applying some pointing model to the offsets in order to compensate for different movements of the telescopes like bending and (thermal) stretching. We would also like to add a wrapper to the PHD2 guiding software, which would allow us to use this well-tested package in addition to our own guiding modules. Furthermore, as mentioned in Section~\ref{sec:scheduling}, a new (non-greedy) scheduler is high up on the wish list.

Using existing software was the goal for pyobs from the beginning. Instead of developing code from scratch it was built on top of widely used Python packages from the astronomical community and beyond. With pyobs-alpaca we already showed that we can bridge towards other protocols, and there probably will be a wrapper for INDI as well, which we can use to add devices for which an INDI driver already exists. There is also a plan to add wrappers for client software like Stellarium\footnote{\url{http://stellarium.org}} or KStars,\footnote{\url{https://edu.kde.org/kstars/}} that will make accessing a remote system easier, e.g.\ for students.

We will continue developing pyobs mainly for our own telescopes, but always trying to be as general as possible, so that it can be used by other observatories. The documentation is a good place to start playing around with pyobs and will be extended continuously. The author of this paper is looking forward to any contribution to pyobs, any comment and suggestion for improvement, and any question via email or GitHub issue tracker.

\section*{Author Contributions}
FVH,  the former PI of MONET, developed a few of the device modules and gave helpful input for the big picture. KR and SM worked on adapting pyobs for the solar telescope. TM implemented the auto-guiding for single stars. SS is the PI of the FTS and responsible for a long list of feature requests and suggestions for improvements.

\section*{Acknowledgments}
The development of pyobs and its modules was only possible by using several Python packages (in alphabetical order):
\begin{itemize}
    \item aiohttp, an asynchronous HTTP Client/Server for asyncio and Python.\footnote{\url{https://docs.aiohttp.org/}}
    \item astroplan, an open source Python package to help astronomers plan observations \citep{astroplan2018}.
    \item Astropy,\footnote{\url{http://www.astropy.org}} a community-developed core Python package for Astronomy \citep{astropy:2013, astropy:2018}.
    \item Astroquery, a set of tools for querying astronomical web forms and databases \citep{2019AJ....157...98G}.
    \item asyncinotify, an async python inotify package.\footnote{\url{https://asyncinotify.readthedocs.io/}}
    \item ccdproc, an Astropy package for image reduction \citep{matt_craig_2017_1069648}.
    \item Cython, an optimising static compiler for the Python programming language.\footnote{\url{https://cython.org}}
    \item lmfit, Non-Linear Least-Squares Minimization and Curve-Fitting for Python.\citep{matt_newville_2021_5570790}
    \item matplotlib, a comprehensive library for creating static, animated, and interactive visualizations in Python.\footnote{\url{https://matplotlib.org}}
    \item numpy, a fundamental package for scientific computing with Python \citep{harris2020array}.
    \item pandas, an open source data analysis and manipulation tool \citep{reback2020pandas,mckinney-proc-scipy-2010}.
    \item Paramiko, a pure-Python implementation of the SSHv2 protocol.\footnote{\url{https://www.paramiko.org/}}
    \item Photutils, an Astropy package for detection and photometry of astronomical sources \citep{larry_bradley_2020_4044744}.
    \item py-expression-eval, a Python mathematical expression evaluator.\footnote{\url{https://github.com/AxiaCore/py-expression-eval/}}
    \item PyQt5, a set of Python bindings for Qt application framework.\footnote{\url{https://www.riverbankcomputing.com/software/pyqt/}}
    \item python-aravis, a Pythonic interface to the auto-generated aravis bindings.\footnote{\url{https://github.com/SintefManufacturing/python-aravis}}
    \item python-daemon, Python library to implement a well-behaved Unix daemon process.\footnote{\url{https://pagure.io/python-daemon/}}
    \item python-telegram-bot, a Python wrapper for using Telegram.\footnote{\url{https://github.com/python-telegram-bot/python-telegram-bot}}
    \item python-zwoasi, a Python binding to the ZWO ASI version 2 library..\footnote{\url{https://github.com/python-zwoasi/python-zwoasi}}
    \item PyYAML, a full-featured YAML framework for the Python programming language.\footnote{\url{https://pyyaml.org}}
    \item qasync, an implementation of the PEP 3156 event-loop to be used in PyQt applications.\footnote{\url{https://github.com/CabbageDevelopment/qasync}}
    \item scipy, a package for fundamental algorithms for scientific computing in Python \citep{2020SciPy-NMeth}.
    \item SEP, a Python and C library for Source Extraction and Photometry \citep{Barbary2016,kyle_barbary_2017_896928}, based on Source Extractor \citep{1996A&AS..117..393B}.
    \item single-source, a single source of truth for version and name of a project.\footnote{\url{https://github.com/rabbit72/single-source}}
    \item Slixmpp, an XMPP library for Python 3.7+.\footnote{\url{https://slixmpp.readthedocs.io/}}148 x 57 x 157 
\end{itemize}

Some more packages are currently used by pyobs but not mentioned here, since they are going to be replaced soon. The GUI uses icons from the "Crystal Clear" set.\footnote{\url{https://commons.wikimedia.org/wiki/Crystal_Clear}}

Running our own instance of the LCO Observation Portal as well as connecting it to pyobs was made possible with the help of the great team at Las Cumbres Observatory (LCO). We also use parts of the frontend of their science archive. Both are parts of the LCO Observatory Control System (OCS).\footnote{\url{https://observatorycontrolsystem.github.io/}}

\begin{table}
  \caption{List of available modules in the core package and in external packages.}
  \label{tbl:modules}
  \centering
  \begin{tabularx}{\textwidth}{llX}
    \multicolumn{3}{c}{\textbf{Core package (\texttt{pyobs.modules.})}} \\
    \hline
    \textbf{Module} & \textbf{Package} & \textbf{Description} \\
    \hline\hline
    \texttt{DummyCamera} & \texttt{camera} & Dummy camera for testing \\
    \texttt{DummySpectrograph} & \texttt{camera} & Dummy spectrograph for testing \\
    \hline
    \texttt{FlatField} & \texttt{flatfield} & Taking a flat-field series \\
    \texttt{FlatFieldPointing} & \texttt{flatfield} & Pointing for flat-fields \\
    \texttt{FlatFieldScheduler} & \texttt{flatfield} & Scheduler for flat-fields \\
    \hline
    \texttt{FocusModel} & \texttt{focus} & Temperature model for focus \\
    \texttt{FocusSeries} & \texttt{focus} & Auto-focus series \\
    \hline
    \texttt{ImageWatcher} & \texttt{image} & Watch directory for new images \\
    \texttt{ImageWriter} & \texttt{image} & Write new images to disk \\    
    \texttt{Seeing} & \texttt{image} & Measure seeing in images \\    
    \hline
    \texttt{AutoGuiding} & \texttt{pointing} & Auto-guiding with external camera \\
    \texttt{Acquisition} & \texttt{pointing} & Fine acquisition \\
    \texttt{ScienceFrameGuiding} & \texttt{pointing} & Auto-guiding with science camera \\
    \texttt{DummyAcquisition} & \texttt{pointing} & Dummy acquisition for testing \\
    \texttt{DummyGuiding} & \texttt{pointing} & Dummy guiding for testing \\
    \hline
    \texttt{Mastermind} & \texttt{robotic} & Main robotic module \\
    \texttt{PointingSeries} & \texttt{robotic} & Automated pointing series \\
    \texttt{Scheduler} & \texttt{robotic} & Task scheduler \\
    \hline
    \texttt{DummyRoof} & \texttt{roof} & Dummy roof for testing \\
    \hline
    \texttt{DummyTelescope} & \texttt{telescope} & Dummy telescope for testing \\
    \hline
    \texttt{AutonomousWarning} & \texttt{utils} & Acoustic warning in robotic mode \\
    \texttt{HttpFileCache} & \texttt{utils} & File cache \\
    \texttt{Kiosk} & \texttt{utils} & Take images and publish on website \\
    \texttt{Telegram} & \texttt{utils} & Telegram interface \\
    \texttt{Trigger} & \texttt{utils} & Event trigger \\
    \hline
    \texttt{Weather} & \texttt{weather} & Connection to pyobs-weather \\
    \hline \\ [0.1 em]
    \multicolumn{3}{c}{\textbf{External packages}} \\
    \hline
    \textbf{Module} & \textbf{Package} & \textbf{Description} \\
    \hline\hline
    \texttt{AlpacaTelescope} & \texttt{pyobs\_alpaca} & Telescope connected via ASCOM Alpaca \\
    \texttt{AlpacaFocuser} & \texttt{pyobs\_alpaca} & Focus unit connected via ASCOM Alpaca \\
    \texttt{AlpacaDome} & \texttt{pyobs\_alpaca} & Dome connected via ASCOM Alpaca \\
    \texttt{AravisCamera} & \texttt{pyobs\_aravis} & Aravis network cameras \\
    \texttt{AsiCamera} & \texttt{pyobs\_asi} & ZWO ASI cameras \\
    \texttt{AsiCoolCamera} & \texttt{pyobs\_asi} & ZWO ASI cameras with active cooling \\
    \texttt{FliCamera} & \texttt{pyobs\_fli} & FLI cameras \\
    \texttt{GUI} & \texttt{pyobs\_gui} & Graphical user interface for remote access \\
    \texttt{Pilar} & \texttt{pyobs\_pilar} & Pilar telescopes \\
    \texttt{SbigCamera} & \texttt{pyobs\_sbig} & SBIG cameras \\
    \texttt{SbigFilterCamera} & \texttt{pyobs\_sbig} & SBIG cameras with filter wheel \\
    \texttt{Sbig6303eCamera} & \texttt{pyobs\_sbig} & SBIG 6303e \\
    \hline
  \end{tabularx}
\end{table}

\bibliographystyle{frontiersinSCNS_ENG_HUMS} 
\bibliography{pyobs}



\end{document}